\documentclass[twocolumn]{aastex63}

\shorttitle{How does the polar dust affect the correlation between CF and $\lambda_{\rm Edd}$ of quasars?}
\shortauthors{Toba et al.}

\begin{document}

\title{How does the Polar Dust affect the Correlation between Dust Covering Factor and Eddington Ratio in Type 1 Quasars Selected from the Sloan Digital Sky Survey Data Release 16?}

\correspondingauthor{Yoshiki Toba}
\email{toba@kusastro.kyoto-u.ac.jp}

\author[0000-0002-3531-7863]{Yoshiki Toba}
\affiliation{Department of Astronomy, Kyoto University, Kitashirakawa-Oiwake-cho, Sakyo-ku, Kyoto 606-8502, Japan}
\affiliation{Academia Sinica Institute of Astronomy and Astrophysics, 11F of Astronomy-Mathematics Building, AS/NTU, No.1, Section 4, Roosevelt Road, Taipei 10617, Taiwan}
\affiliation{Research Center for Space and Cosmic Evolution, Ehime University, 2-5 Bunkyo-cho, Matsuyama, Ehime 790-8577, Japan}

\author[0000-0001-7821-6715]{Yoshihiro Ueda}
\affiliation{Department of Astronomy, Kyoto University, Kitashirakawa-Oiwake-cho, Sakyo-ku, Kyoto 606-8502, Japan}

\author[0000-0003-3105-2615]{Poshak Gandhi}
\affiliation{Department of Physics and Astronomy, University of Southampton, Highfield, Southampton SO17 1BJ}

\author[0000-0001-5231-2645]{Claudio Ricci}
\affiliation{N\'ucleo de Astronom\'{\i}a de la Facultad de Ingenier\'{\i}a, Universidad Diego Portales, Av. Ej\'ercito Libertador 441, Santiago, Chile}
\affiliation{Kavli Institute for Astronomy and Astrophysics, Peking University, Beijing 100871, People's Republic of China}

\author[0000-0002-4193-2539]{Denis Burgarella}
\affiliation{Aix Marseille Univ. CNRS, CNES, LAM Marseille, France}

\author[0000-0003-3441-903X]{Veronique Buat}
\affiliation{Aix Marseille Univ. CNRS, CNES, LAM Marseille, France}

\author[0000-0002-7402-5441]{Tohru Nagao}
\affiliation{Research Center for Space and Cosmic Evolution, Ehime University, 2-5 Bunkyo-cho, Matsuyama, Ehime 790-8577, Japan}

\author{Shinki Oyabu}
\affiliation{Institute of Liberal Arts and Sciences, Tokushima University, Minami Jousanjima-Machi 1-1, Tokushima, Tokushima 770-8502, Japan}

\author{Hideo Matsuhara}
\affiliation{Institute of Space and Astronautical Science, Japan Aerospace Exploration Agency, 3-1-1 Yoshinodai, Chuo-ku, Sagamihara, Kanagawa 252-5210, Japan}
\affiliation{Department of Space and Astronautical Science,The Graduate University for Advanced Studies, SOKENDAI, 3-1-1 Yoshinodai,Chuo-ku, Sagamihara, Kanagawa 252-5210, Japan}

\author[0000-0001-5615-4904]{Bau-Ching Hsieh}
\affiliation{Academia Sinica Institute of Astronomy and Astrophysics, 11F of Astronomy-Mathematics Building, AS/NTU, No.1, Section 4, Roosevelt Road, Taipei 10617, Taiwan}



\begin{abstract}

We revisit the dependence of covering factor (CF) of dust torus on physical properties of active galactic nuclei (AGNs) by taking into account an AGN  polar dust emission.
The CF is converted from a ratio of infrared (IR) luminosity contributed from AGN dust torus ($L_{\rm IR}^{\rm torus}$) and AGN bolometric luminosity ($L_{\rm bol}$), by assuming a non-linear relation between luminosity ratio and intrinsic CF.
We select 37,181 type 1 quasars at $z < 0.7$ from the Sloan Digital Sky Survey Data Release 16 quasar catalog.
Their $L_{\rm bol}$, black hole mass ($M_{\rm BH}$), and Eddington ratio ($\lambda_{\rm Edd}$) are derived by spectral fitting with {\tt QSFit}.
We conduct spectral energy distribution decomposition by using {\tt X-CIGALE} with clumpy torus and polar dust model to estimate $L_{\rm IR}^{\rm torus}$ without being affected by the contribution of stellar and AGN polar dust to IR emission.
For 5720 quasars whose physical quantities are securely determined, we perform a correlation analysis on CF and (i) $L_{\rm bol}$, (ii) $M_{\rm BH}$, and (iii) $\lambda_{\rm Edd}$.
As a result, anti-correlations for CF--$L_{\rm bol}$, CF--$M_{\rm BH}$, and CF--$\lambda_{\rm Edd}$ are confirmed.
We find that incorporating the AGN polar dust emission makes those anti-correlations stronger which are compared to those without considering it.
This indicates that polar dust wind provably driven by AGN radiative pressure is one of the key components to regulate obscuring material of AGNs.
\end{abstract}



\section{Introduction} 
\label{intro}
An obscuring dusty structure surrounding a supermassive black hole (SMBH) plays an important role in the diversity of observational phenomena of active galactic nuclei (AGNs) in the context of unified model \citep[e.g.,][]{Antonucci,Urry}.
The obscuring material that is often called ``dust torus'' \citep[e.g.,][]{Krolik} is expected to be compact ($<$ 10 pc) and responsible for infrared (IR) emission from AGNs \citep[e.g.,][]{Rees,Jaffe,Tristram07}.

Although it has been believed that dust torus is a key component of the AGNs, many challenges still remain; 
What is the structure of the circumnuclear dust?
What is the main physical mechanism to regulate the obscuring material surrounding AGNs?
These have been extensively studied both from theoretical side \citep[e.g.,][]{Wada02,Schartmann,Fritz,Kawakatu08,Nenkova08,Nenkova08b,Wada,Wada15,Siebenmorgen,Stalevski,Namekata,Tanimoto} and observational side \citep[e.g.,][]{Suganuma,Ueda07,Gandhi09,Gandhi15,Alonso,Kishimoto,Ramos,Ricci14,Brightman,Imanishi16,Imanishi18,Baba,Balokovic,Izumi,Honig,Marchesi} \citep[see also][and references therein]{Netzer,Ramos17,Hickox}.

We focus here on the geometrical covering fraction of dust torus, a fundamental parameter in the AGN unified model.
The dust covering factor (CF) is defined as the fraction of the sky, as seen from the AGN center, that is blocked by heavily obscuring dust.
The CF also provides important information on the number of obscured AGNs.
Many works have reported that the CF may depend on AGN luminosity wherein CF is often defined as type 2 AGN fraction \citep[e.g.,][]{Simpson,Toba_12,Toba_13,Toba_14,Toba_17}, the IR-to-bolometric luminosity \citep[e.g.,][]{Maiolino,Gandhi09,Gu,Ma,Roseboom}, or X-ray obscured AGN fraction \citep[e.g.,][]{Ueda,Ueda14,Hasinger,Burlon,Merloni} (but see also e.g., \citealt{Dwelly,Lawrence10,Mateos17,Ichikawa} who reported weak or no significant dependence).
The CF may also depend on redshift \citep[e.g.,][]{LaFranca,Hasinger,Gu,Merloni,Ueda14} as well as counterargument of no significant evolution at least a certain redshift range \citep[e.g.,][]{Toba_14,Vito}.

Recently, \cite{Ricci} reported that the key parameter determining X-ray CF may be the Eddington ratio ($\lambda_{\rm Edd}$) rather than AGN luminosity (e.g., bolometric luminosity, $L_{\rm bol}$) based on 731 hard X-ray selected AGNs with a median redshift of 0.0367.
\cite{Ezhikode} performed a correlation analysis for X-ray/optically selected 51 type 1 AGNs at $z < 0.4$, and combining the correlation analysis with simulations, they found that CF is more strongly anti-correlated with $\lambda_{\rm Edd}$ than with $L_{\rm bol}$.
\cite{Zhuang} also reported that CF decreases with increasing $\lambda_{\rm Edd}$ up to $\sim$0.5 based on 76 Palomar-Green (PG) quasars at $z < 0.5$.
\cite{Toba_19a} found that luminosity ratio of 6 $\micron$ and absorption-corrected hard X-ray luminosity for AGNs (that is expected to be an indicator of CF) depends on $\lambda_{\rm Edd}$, and that correlation would be applicable even for a Compton-thick AGN \citep{Toba_20a}.

On the other hand, \cite{Cao} reported no correlation between the near-IR (NIR)-to-bolometric luminosity ratio (that is an indicator of CF) and $\lambda_{\rm Edd}$ for PG quasars, whereas there is a significant correlation between CF--$L_{\rm bol}$ and CF--the central black hole mass ($M_{\rm BH}$).
These trends were also reported by \cite{Ma} based on 17,639 quasars at $0.76 < z < 1.17$ selected from the Sloan Digital Sky Survey \citep[SDSS: ][]{York}.

The discrepancy in above works may be partially because the difference in sample selection, redshift distribution, and the definition of CF.
In particular, statical works on CF for quasars tend to employ $L_{\rm IR}$/$L_{\rm bol}$ as CF.
However, recent works suggested that $L_{\rm IR}$/$L_{\rm bol}$ may not be always a good indicator of CF caused by  anisotropy in the IR emission of dust torus \citep[e.g.,][]{Stalevski,Mateos17,Zhuang}.
Because $L_{\rm IR}$/$L_{\rm bol}$ and CF are not connected by a simple linear relation, one needs a conversion from the luminosity ratio to the CF (see Section \ref{CF}).
In addition, the derived NIR luminosity for CF may be contaminated by stellar emission, which may not be negligible especially for less luminous quasars \citep[e.g.,][]{Mateos,Toba_19a}.
Furthermore, more recently, polar dust wind probably driven by AGN radiative pressure has been shown to be an ubiquitous contributor to the IR emission \citep[e.g.,][]{Honig13,Tristram,Lopez,Leftley,Asmus,Stalevski19}.
\cite{Lyu} demonstrated that the observed variety of their broad-band IR spectral energy distribution (SED) can be explained by considering polar dust component.
As \cite{Asmus} pointed out, we may need to take into account the polar dust emission when deriving CF from $L_{\rm IR}$/$L_{\rm bol}$.

In this work, we revisit the relationship between CF and $L_{\rm bol}$, $M_{\rm BH}$, and $\lambda_{\rm Edd}$ for type 1 quasars at $z < 0.7$ selected from the SDSS quasar catalog Data Release (DR) 16 \citep[DR16Q; ][]{Lyke} that has been recently published.
For this enormous number of type 1 quasars, we extract IR emission purely from dust torus after correcting for the stellar and polar dust emission based on the SED fitting with a clumpy torus model and polar dust model, and convert the luminosity ratio to intrinsic CF following \cite{Stalevski}.
This paper is organized as follows.
Section \ref{DA} describes the sample selection of type 1 quasars, our SED modeling, and spectral fitting to the SDSS spectra.
In Section \ref{R}, we present the results of the SED fitting, spectral fitting, and the dependence of CF on $L_{\rm bol}$, $M_{\rm BH}$, and $\lambda_{\rm Edd}$.
In Section \ref{D}, we show a correlation analysis, and discuss possible uncertainties of the results and comparison of previous works.
We summarize the results of the study in Section \ref{Sum}.
Throughout this paper, the adopted cosmology is a flat universe with $H_0$ = 70.0 km s$^{-1}$ Mpc$^{-1}$, $\Omega_M$ = 0.30 and $\Omega_{\Lambda}$ = 0.70.
Unless otherwise noted, $z$ refers to the spectroscopic redshift and an initial mass function (IMF) of  \cite{Chabrier} is assumed.

\section{Data and Analysis} 
\label{DA}

\subsection{Sample Selection} 
\label{SS}

\begin{figure}
\centering
\includegraphics[width=0.4\textwidth]{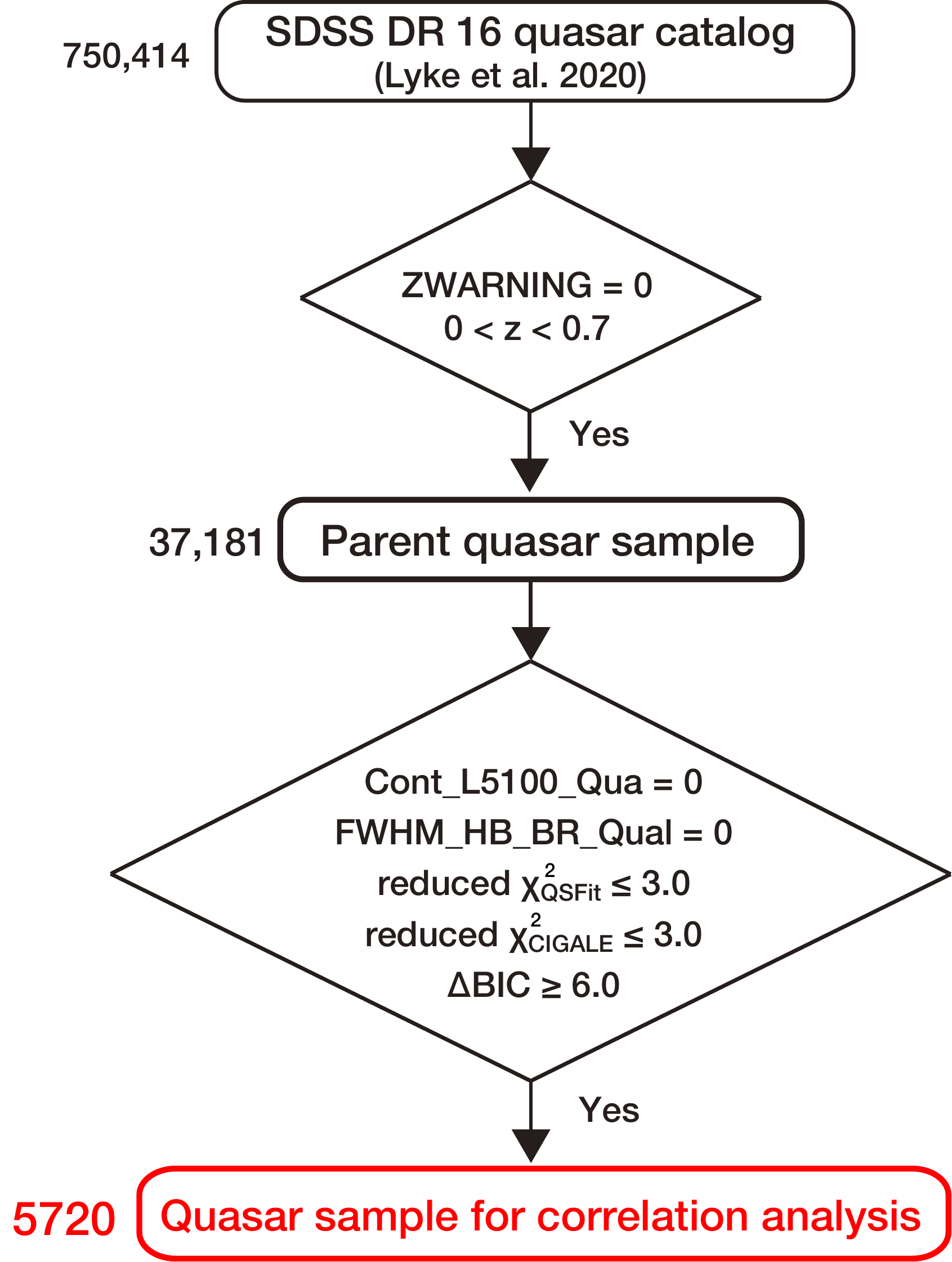}
\caption{Flow chart of the process to quasar sample for a correlation analysis.}
\label{FC}
\end{figure}

A flowchart of our sample selection process is shown in Figure \ref{FC}.
In this work, we focus on spectroscopically confirmed type 1 quasars whose $L_{\rm bol}$, $M_{\rm BH}$ and $\lambda_{\rm Edd}$ can be securely estimated.
We sampled quasars that were drawn from the DR16Q (v4\footnote{\url{https://data.sdss.org/sas/dr16/eboss/qso/DR16Q/}}) that contains 750,414 type 1 quasars up to $z \sim 6.0$.

Aside from the SDSS imaging data, 3.4 and 4.6 $\micron$ data taken from the Wide-field Infrared Survey Explorer \citep[WISE;][]{Wright} and information on source variability from the Palomar Transient Factory \citep[PTF;][]{Law,Rau} were used for target selection of quasars \citep[see Sections 2.1 and 2.2 in ][for more detail]{Lyke}.
The targeted objects were automatically classified as {\tt QSO}, {\tt STAR}, or {\tt GALAXY} based on the SDSS spectra.
In addition, objects that were initially classified as {\tt QSO} at $z_{\rm pipe} > 3.5$ by the SDSS pipeline were reclassified for visual inspection \citep[see Section 3.2 in][]{Lyke}.

Compared to previous SDSS quasar catalogs (e.g., DR7Q; \citealt{Schneider}, DR9Q; \citealt{Paris_12}, and DR14Q; \citealt{Paris}), DR16Q extends coverage of luminosity to much less luminous quasars given a redshift (see Figures 7 in \citealt{Paris} and \citealt{Lyke}).
Hence our quasar sample is expected to cover a wide range of $L_{\rm bol}$, $M_{\rm BH}$, and $\lambda_{\rm Edd}$  compared with those derived from previous SDSS quasar catalogs employed in previous works on CF \citep[e.g.,][]{Gu,Ma,Roseboom}.

We first narrowed down the sample to objects with {\tt ZWARNING = 0}, which ensures a confident spectroscopic classification and redshift measurement for quasars \citep{Bolton}.
We then extracted objects with $0 < z < 0.7$ because we aim to estimate $M_{\rm BH}$ based on a recipe using H$\beta$ (see Section \ref{qsfit}) that is detectable for objects at $z$ up to $\sim 0.7$ given a spectral coverage of SDSS spectrograph.
Note that it is possible to measure $M_{\rm BH}$ for objects at $z > 0.7$ based on recipes using other emission lines (such as C{\,\sc iv} and Mg{\,\sc ii}).
But, to avoid being affected by possible systematic uncertainties of $M_{\rm BH}$ because of using different recipes, we adopted the same recipe for all quasars and perform a correlation analysis (see Sections \ref{qsfit} and  \ref{CA}).
Eventually, 37,181 quasars were selected as a sample.
We note that the number of quasar sample at $z < 0.7$ selected from the DR16Q is significantly increased by $\sim$15,000 from DR14Q.

\subsection{Photometric data}  
\label{pd}

For those 37,181 quasars, we compiled optical to mid-IR (MIR) photometry, most of which are already available in DR16Q.
In addition to SDSS optical data ($u$, $g$, $r$, $i$, and $z$) corrected for Galactic extinction \citep{Schlafly}, we utilized the IR data taken with Two Micron All Sky Survey \citep[2MASS;][]{Skrutskie}, the UKIRT Infrared Deep Sky Survey \citep[UKIDSS;][]{Lawrence}, and the Wide-field Infrared Survey Explorer \citep[WISE;][]{Wright} (see Section 7 in \citealt{Lyke} for a complete description of the above data).

Note that DR16Q newly employed unWISE \citep{Lang} instead of ALLWISE catalog \citep{Cutri}.
unWISE provides deep 3.4 and 4.6 $\micron$ data that are force-photometered at the locations of SDSS sources \citep{Lang16}.
On the other hand, WISE 12 and 22 $\micron$ data are still quite useful for this work because those wavelengths are expected to trace the emission from the dusty torus \citep[e.g.,][]{Toba_14}.
Hence we compiled those MIR data from ALLWISE with a search radius of 2$\arcsec$ in the same manner as \cite{Paris}.

Consequently, we have at maximum 13 photometric data ($u$, $g$, $r$, $i$, $z$, $Y$, $J$, $H$, $K$/$K_{\rm s}$, and 3.4, 4.6, 12, and 22 $\micron$) for the SED fitting (see Section \ref{cigale}).
Among 37,181 quasars, 3345 (9.0 \%), 5515 (14.8 \%), 34,683 (93.3 \%), and 37,155 (99.9 \%) objects are detected by 2MASS, UKIDSS, ALLWISE, unWISE, respectively.
We used profile-fit photometry for 2MASS-detected sources with {\sf rd\_flg}=2\footnote{\url{https://old.ipac.caltech.edu/2mass/releases/allsky/doc/sec4_4d.html}} and that for WISE 12 and/or 22 $\micron$-detected sources with {\sf cc\_flag}=0\footnote{\url{http://wise2.ipac.caltech.edu/docs/release/allsky/expsup/sec4_4g.html}} (which ensures clean photometry without being affected by possible artifacts) at each band.
If an object lies outside the UKIDSS footprint, its NIR flux densities are taken from 2MASS (if that object is detected).
Otherwise, we always refer to the UKIDSS as NIR data.

\subsection{Broad-band SED Fitting with {\tt X-CIGALE}}  
\label{cigale}
To derive IR luminosity contributed only from AGN dust torus ($L_{\rm IR}^{\rm torus}$), we conducted the SED fitting by considering the energy balance between the UV/optical and IR.
We employed a new version of Code Investigating GALaxy Emission \citep[{\tt CIGALE}; ][]{Burgarella,Noll,Boquien}
 so-called {\tt X-CIGALE\footnote{\url{https://gitlab.lam.fr/gyang/cigale/tree/xray}}} \citep{Yang}.
This code enables us to handle many parameters, such as star formation history (SFH), single stellar population (SSP), attenuation law, AGN emission, dust emission, and radio synchrotron emission \cite[see e.g.,][]{Boquien14,Buat_14,Buat,Boquien16,Ciesla_17,LoFaro,Toba_19b,Burgarella20}. 
Aside from an implementation of the SED fitting even for X-ray data, {\tt X-CIGALE} incorporates a clumpy two-phase torus model \citep[{\tt SKIRTOR}\footnote{\url{https://sites.google.com/site/skirtorus}};][]{Stalevski12,Stalevski} and a polar dust emission as AGN templates.
According to \cite{Yang}, we consider the dust responsible for type 1 AGN obscuration as polar dust (i.e.., the polar dust provides obscuration for the nucleus and the additional IR emission from the absorbed energy, see Figure 4 in \citealt{Yang} for a schematic view of polar dust).
Parameter ranges used in the SED fitting are tabulated in Table \ref{Param}.

\begin{table}
\renewcommand{\thetable}{\arabic{table}}
\centering
\caption{Parameter Ranges Used in the SED Fitting with {\tt X-CIGALE}} 
\label{Param}
\begin{tabular}{lc}
\tablewidth{0pt}
\hline
\hline
Parameter & Value\\
\hline
\multicolumn2c{Delayed SFH}\\
\hline
$\tau_{\rm main}$ (Gyr) 	& 0.5, 1.0, 2.0, 4.0, 6.0, 8.0 \\
age  (Gyr) 					& 5.0\\
\hline
\multicolumn2c{SSP \citep{Bruzual}}\\
\hline
IMF				&	\cite{Chabrier} \\
Metallicity		&	0.02 \\
\hline
\multicolumn2c{Dust Attenuation \citep{CF00}}\\
\hline
$A_{\rm V}^{\rm ISM}$ 	& 0.01, 0.1, 0.2, 0.3, 0.4, 0.6, 0.8, 1.0 \\ 
 slope\_ISM & -0.7\\
 slope\_BC & -1.3 \\
\hline
\multicolumn2c{AGN Disk +Torus Emission \citep{Stalevski}}\\
\hline
$\tau_{\rm 9.7}$ 			&  	3, 5, 9 		\\
$p$							&	0.0, 1.0, 1.5	\\
$q$							&	0.0, 1.0, 1.5	\\
$\Delta$ (\arcdeg)			&	10, 30, 50		\\
$R_{\rm max}/R_{\rm min}$ 	& 	30 				\\
$\theta$ (\arcdeg)			&	0				\\
$f_{\rm AGN}$ 				& 	0.5, 0.6, 0.7, 0.8, 0.9, 0.99\\
\hline
\multicolumn2c{AGN Polar Dust Emission \citep{Yang}}\\
\hline
extinction low					&	SMC		\\
$E(B-V)$						&	0.0, 0.05, 0.1, 0.15, 0.2, 0.3, 0.4	\\
$T_{\rm dust}^{\rm polar}$ (K)	&	100.0, 500.0, 1000.0	\\
Emissivity $\beta$				&	1.6\\
\hline
\end{tabular}
\end{table}
We adopted a delayed SFH model, assuming a single starburst with an exponential decay \citep[e.g.,][]{Ciesla_15,Ciesla_16}, where we fixed the age of the main stellar population in the galaxy that is the same as what we used for galaxy template when fitting to the SDSS spectra (see Section \ref{qsfit}).
We parameterized $e$-folding time of the main stellar population ($\tau_{\rm main}$). 
The influence of fixing the age of the main stellar population on CF is discussed in Section \ref{age}.

We chose the SSP model \citep{Bruzual}, assuming the IMF of \cite{Chabrier}, and the standard nebular emission model included in {\tt X-CIGALE} \citep[see ][]{Inoue}.

For attenuation of dust associated with host galaxy, we utilized a model provided by \cite{CF00} that has two different power-law attenuation curves; the power law slope of the attenuation in the interstellar medium (ISM) and birth clouds (BC) that were fixed in this work.
We parameterized the $V$-band attenuation in the ISM ($A_{\rm V}^{\rm ISM}$).

AGN emission from accretion disk and dust torus was modeled by using the {\tt SKIRTOR}, a clumpy two-phase torus model produced in the framework of the 3D radiative-transfer code, {\tt SKIRT} \citep{Baes,Camps}.
This torus model consists of 7 parameters; torus optical depth at 9.7 $\micron$ ($\tau_{\rm 9.7}$), torus density radial parameter ($p$), torus density angular parameter ($q$), angle between the equatorial plane and edge of the torus ($\Delta$), ratio of the maximum to minimum radii of the torus ($R_{\rm max}/R_{\rm min}$), the viewing angle ($\theta$), and the AGN fraction in total IR luminosity ($f_{\rm AGN}$). 
In order to avoid a degeneracy of AGN templates \citep[see ][]{Yang} and to estimate a reliable CF (see Section \ref{CF}), we fixed $R_{\rm max}/R_{\rm min}$ and $\theta$ that are optimized for type 1 quasars.
We note that $R_{\rm max}/R_{\rm min}$ and $\theta$ are insensitive to the CF at least for type 1 AGNs as \cite{Stalevski12} reported.

In {\tt X-CIGALE}, the polar dust emission is implemented as a gray body emission with an empirical extinction curve \citep[see Section 2.4 in][for more detail]{Yang}.
The gray body is formulated as $1-e^{\tau (\nu)}$ $\nu^{\beta}\,B_{\rm \nu} (T_{\rm dust}^{\rm polar})$, where $\nu$ is the frequency, $\beta$ is the emissivity index of the dust, and $B_{\rm \nu} (T_{\rm dust}^{\rm polar})$ is the Planck function.
$\tau \equiv (\nu/\nu_0)^{\beta}$ is the optical depth. 
$\nu_0$ is the frequency where optical depth equals unity \citep{Draine}.
$\nu_0$ = 1.5 THz ($\lambda_0$ = 200 $\micron$) in {\tt X-CIGALE} and $\beta$ is fixed to be 1.6 \citep[e.g.,][]{Casey}, and thus we parameterized $T_{\rm dust}^{\rm polar}$ in the same manner as \cite{Toba_20b}.
We adopted the SMC extinction curve \citep{Prevot} that may be reasonable for AGN dust \citep[e.g.,][]{Pitman,Kuhn,Salvato}, in which we parameterized $E(B-V)$.
$E(B-V)$ = 0 means that polar dust component is not necessary.
We confirmed that our conclusion is not significantly affected if we assumed another extinction curve.

In the framework of our SED modeling, IR luminosity can be described by a combination of stellar component ($L_{\rm IR}^{\rm stellar}$), AGN accretion disk component ($L_{\rm IR}^{\rm disk}$), AGN dust torus component ($L_{\rm IR}^{\rm torus}$), and AGN polar dust component ($L_{\rm IR}^{\rm polar}$), which enables to estimate $L_{\rm IR}^{\rm torus}$ that is relevant to CF through the SED decomposition.

Under the parameter setting described in Table \ref{Param}, we fit the stellar and AGN (accretion disk, dust torus, and polar dust) components to at most 13 photometric data as described in Section \ref{pd}.
Although the photometry employed in DR16Q is not identical, the profile-fit photometry is used for the SDSS and 2MASS \citep{Lyke}, while the UKIDSS and unWISE data are forced-photometry at the SDSS centroids \citep{Aihara,Lang}.
Therefore, the flux densities in the optical-MIR bands in DR16Q are expected to trace the total flux densities, suggesting that the influence of different photometry is likely to be small (see also Section \ref{mock}).
We used flux density in a band when the signal-to-noise ratio (S/N) is greater than 3 at that band.
Following \cite{Toba_19a}, we put 3$\sigma$ upper limits for objects with {\sf ph\_qual}=`U' (which means upper limit on magnitude) in the MIR-bands.

\subsection{Optical Spectral Fitting with {\tt QSFit}}  
\label{qsfit}

In order to derive $M_{\rm BH}$, $L_{\rm bol}$, and $\lambda_{\rm Edd}$ of our quasar sample, we conducted a spectral fitting to SDSS spectra by using the Quasar Spectral Fitting package \citep[{\tt QSFit} v1.3.0;][]{Calderone}\footnote{\url{https://qsfit.inaf.it}}.
This code fits optical spectra by taking into account (i) AGN continuum with a single power law, (ii) Balmer continuum modeled by \cite{Grandi} and \cite{Dietrich}, (iii) host galaxy component with an empirical SED template, (iv) iron blended emission lines with UV-optical templates \citep{Veron,Vestergaard01}, and (v) emissions line with  Gaussian components.
{\tt QSFit} fits all the components simultaneously following a Levenberg-Marquardt least-squares minimization algorithm with {\tt MPFIT} \citep{Markwardt} procedure.

The main purpose of this spectral fitting is to measure FWHM of H$\beta$ broad component ($FWHM_{\rm H\beta}$) and continuum luminosity at 5100\AA\,($L_{\rm 5100}$) that are ingredients for $M_{\rm BH}$ estimates. 
For the host galaxy contribution\footnote{Since {\tt QSFit} allows to fit the host galaxy component for objects at $z < 0.8$ by default (see Section 2.4 in \citealt{Calderone}), our spectral fitting to the quasar sample (at $z < 0.7$) always takes into account the contribution of host galaxy to the continuum.}, we employed a simulated 5 Gyr old elliptical galaxy template \citep{Silva,Polletta} with allowing the normalization to vary.
After correcting the Galactic extinction provided by \cite{Schlafly} that is same as what we adopted for the SDSS photometry (see Section \ref{pd}), spectral fitting was executed.
Following \cite{Calderone}, we fit the H$\beta$ line with narrow and broad components in which the FWHM of narrow and broad components is constrained in the range of 100 to 1000 and 900 to 15,000 km s$^{-1}$, respectively to allow a good decomposition of the line profile.

Based on outputs from {\tt QSFit}, we estimated $M_{\rm BH}$ with a single-epoch method reported by \cite{Vestergaard};
\begin{eqnarray}
\label{Eq}
\log\left(\frac{M_{\rm BH}}{M_{\sun}}\right) & = & \log\left[\left(\frac{FWHM_{\rm H\beta}}{1000 \,{\rm km \, s^{-1}}} \right)^2 \left(\frac{L_{\rm 5100}}{10^{44} \,{\rm erg \, s^{-1}}}\right)^{0.50}\right] \nonumber \\
& & + (6.91 \pm 0.02).
\end{eqnarray}
$L_{\rm bol}$ was converted from $BC_{\rm 5100} \times L_{\rm 5100}$ where $BC_{\rm 5100}$ = $8.1 \pm  0.4$ is the bolometric correction \citep{Runnoe}.
The uncertainty in $M_{\rm BH}$ is calculated through error propagation of Equation \ref{Eq} while the uncertainty in $L_{\rm bol}$ is propagated from 1$\sigma$ errors of $L_{\rm 5100}$ and $BC_{\rm 5100}$.

\section{Results} 
\label{R}

Here we present physical properties of 37,181 quasars that are derived from the SED fitting with {\tt X-CIGALE} and spectral fitting with {\tt QSFit}, which are summarized in Table \ref{catalog}.
\begin{figure*}
    \centering
    \includegraphics[width=\textwidth]{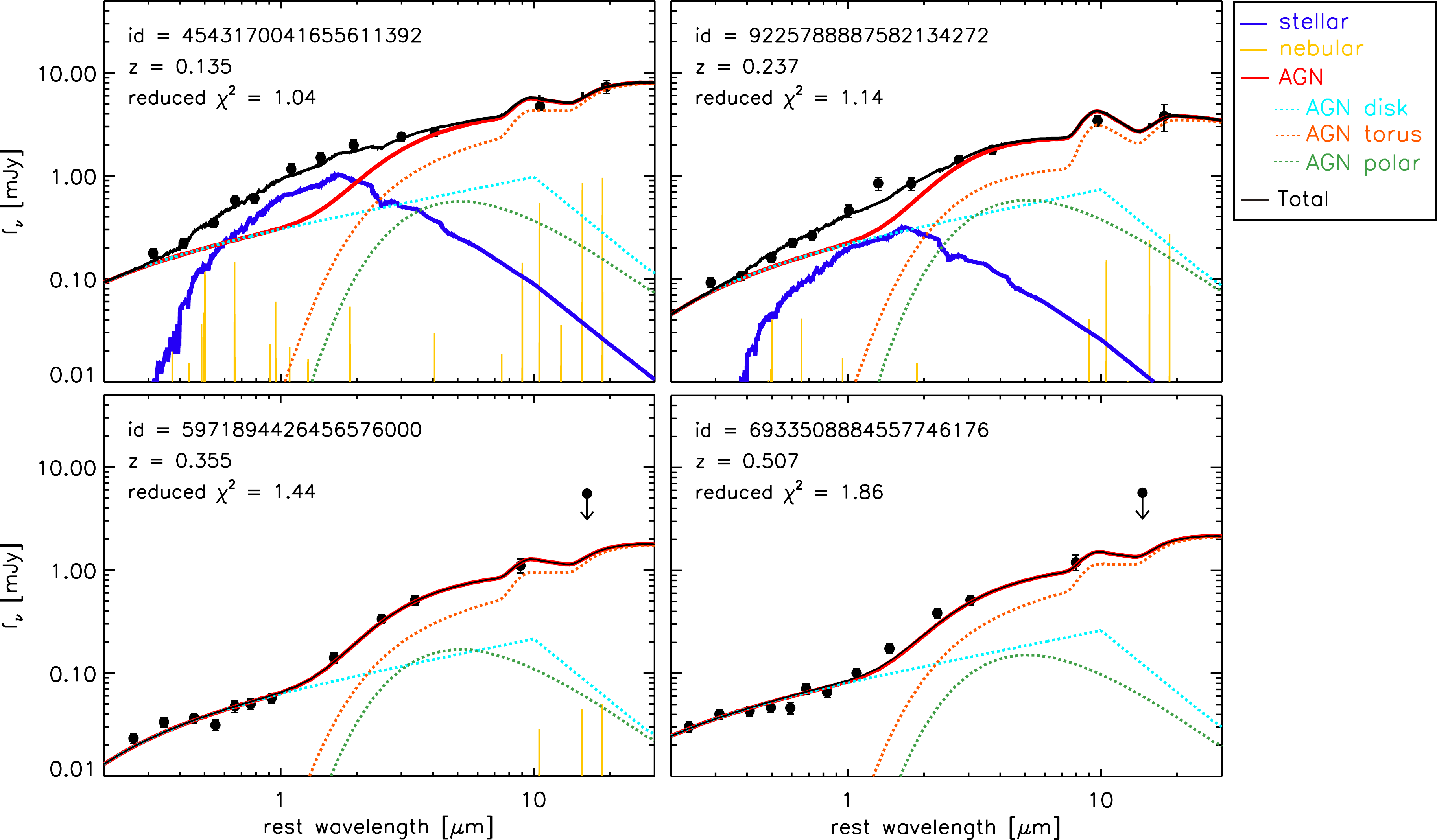}
\caption{Examples of the SED fitting with {\tt X-CIGALE}. The black points are photometric data. The contribution from the stellar, nebular, and AGN components to the total SED are shown as solid blue, yellow, and red lines, respectively. The contribution from AGN accretion disk, dust torus, and polar dust components to the total AGN emission are plotted as dashed cyan, orange, and green lines, respectively. The black solid lines represent the resultant best-fit SED.}
\label{SED}
\end{figure*}

\subsection{Result of the SED Fitting and Bayesian Information Criterion}   
\label{BIC}

Figure \ref{SED} shows examples of the SED fitting with {\tt X-CIGALE}.
We confirm that 34,541/37,181 ($\sim$92.9 \%) objects have reduced $\chi_{\rm CIGALE}^{2} < 3.0$,
meaning that the data are well fitted with the combination of the stellar, nebular, AGN accretion disk, dust torus, and polar dust components by {\tt X-CIGALE}.

On the other hand, it is worth investigating whether polar dust component (that is the main topic in this work) is practically needed to improve the SED fitting.
In order to test the requirement to add an AGN polar dust component to the SED fitting, we calculate the Bayesian information criterion \citep[BIC; ][]{Schwarz} for two fits that are derived with and without polar dust component.
The BIC is defined as BIC = $\chi^{2}$ + $k$ $\times$ ln($n$), where $\chi^{2}$ is non-reduced chi-square, $k$ is the number of degrees of freedom (DOF), and $n$ is the number of photometric data points used for the fitting.
We then compare the results of two SED fittings without/with polar dust component by using $\Delta$BIC = BIC$_{\rm wopolar}$ -- BIC$_{\rm wpolar}$.
The resultant $\Delta$BIC tells whether or not the AGN polar dust component is required to give a better fit with taking into account the difference in DOF \citep[e.g.,][]{Ciesla_18,Buat_19,Aufort,Toba_20c}.
\begin{figure}
\vspace{0.5\baselineskip}
    \centering
    \includegraphics[width=0.4\textwidth]{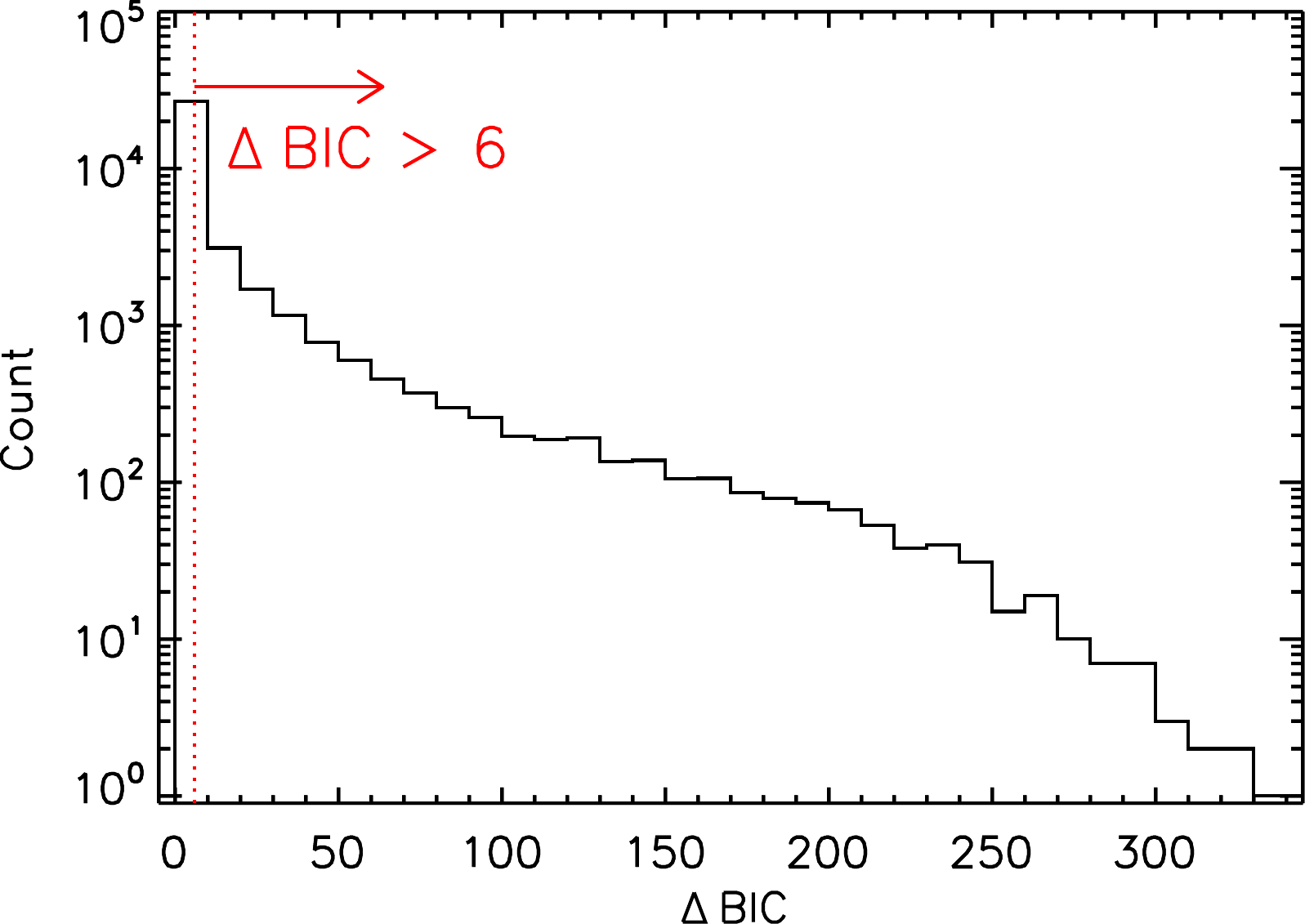}
\caption{Histogram of $\Delta$BIC = BIC$_{\rm wopolar}$--BIC$_{\rm wpolar}$ for the quasar sample. The red dotted line corresponds to $\Delta$BIC = 6 that is a threshold to consider the differences in two fits with and without adding polar dust component.}
\label{fig_BIC}
\end{figure}
\begin{figure*}
    \centering
    \includegraphics[width=0.9\textwidth]{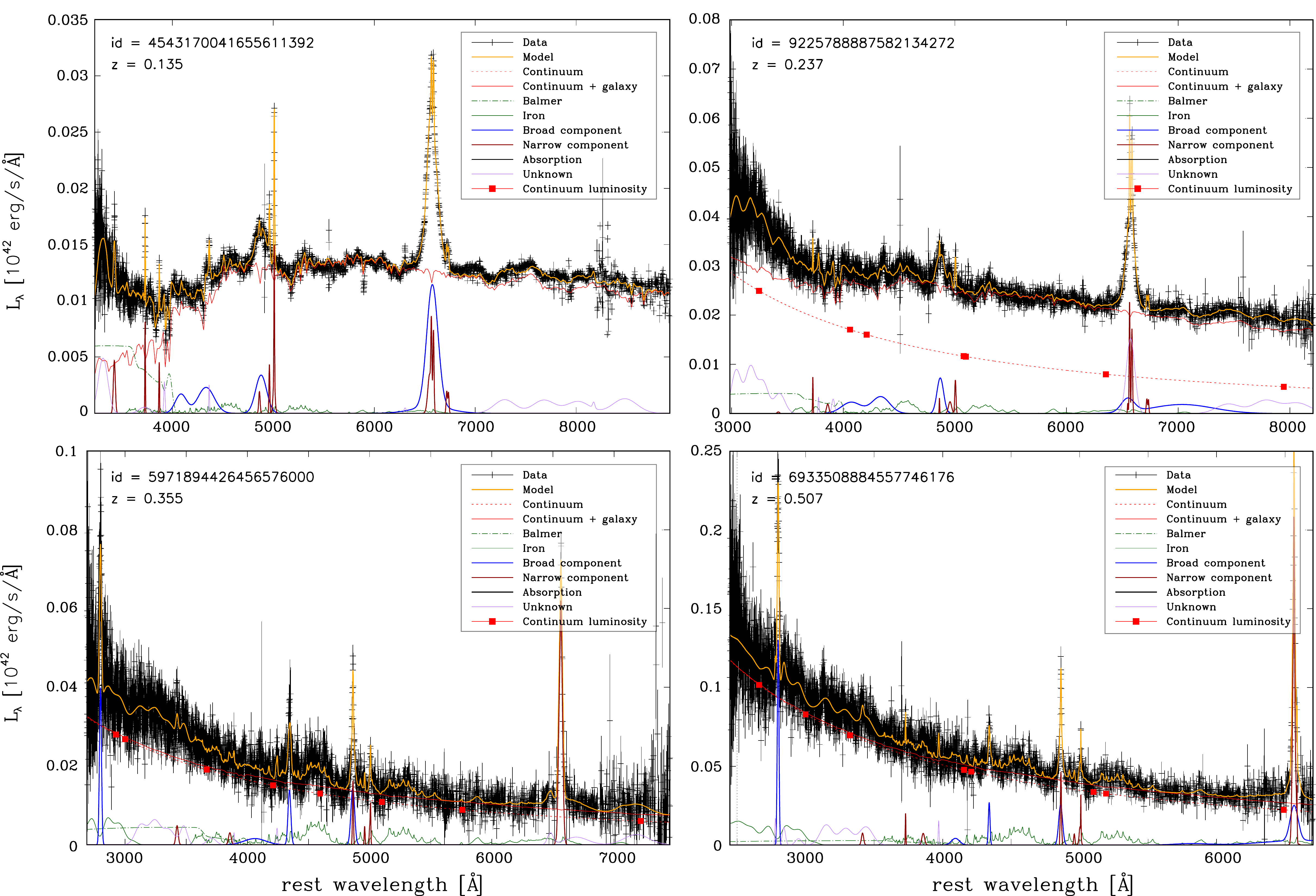}
\caption{Examples of optical spectral fitting with {\tt QSFit}. The same objects plotted in Figure \ref{SED} are presented. The black crosses are observed data. The power-law continuum is shown in dotted red lines while the power-law continuum plus AGN host components are shown in solid red lines. The green dash-dotted and solid curves denote the Balmer continuum and iron continuum, respectively. The brown and blue solid curves show the narrow and broad components of emission lines, respectively. The best-fit spectrum is shown in yellow lines.}
\label{optspec}
\end{figure*}
Figure \ref{fig_BIC} shows the histogram of $\Delta$BIC for our quasar sample.
We adopt $\Delta$BIC = 6  as a threshold to consider the differences in two fits in the same manner as \cite{Buat_19}.
If $\Delta$BIC $>$ 6, this indicates that adding the AGN polar dust component significantly improves the fit.
We find that about 30.0\% of objects satisfy $\Delta$BIC $> 6$. 
This suggests that for a faction of objects in our SDSS quasar sample at $z < 0.7$, an AGN polar dust component may not be necessary.
We note that the threshold we adopted in this work is conservative compared to what originally suggested to be $\Delta$BIC = 2 \citep{Liddle,Stanley}.
Nevertheless, the above result could suggest that our SED modeling may not be enough to constrain of the AGN polar dust emission given a limited number of photometric data in MIR regime, which we should keep in mind for the following analysis.
We remind that the main purpose of this work is to see how the AGN polar dust would affect the CF of type 1 quasars.
Hence, we consider objects with $\chi_{\rm CIGALE}^{2} < 3.0$ and $\Delta$BIC $>$ 6 for the correlation analysis (see Section \ref{CF}).

\subsection{Result of the Spectral Fitting}   
\label{SpF}

Figure \ref{optspec} shows examples of the optical spectral fitting with {\tt QSFit}.
We confirm that 36,855/37,181 ($\sim$99.1\%) objects have reduced $\chi_{\rm QSFit}^{2} < 3.0$ while only 170 ($\sim$0.5 \%) objects are failed to fit due to low S/N. 
This indicates that the SDSS spectra of our sample are well-fitted by {\tt QSFit}.
Note that objects showing top panel in Figure \ref{optspec} are expected to have large contribution from host galaxy to optical spectrum according to the SED fitting (see top panel in Figure \ref{SED}).
We confirm this is the case for their optical spectra, suggesting that our SED modeling and spectral fitting give consistent results.

In addition to goodness of fitting (i.e., $\chi_{\rm QSFit}^{2}$), {\tt QSFit} provides ``quality flags'' for measurements to assess the reliability of the results \citep[see Appendix C in][for more detail]{Calderone}.
We considered {\tt Cont\_5100\_Qual} and {\tt HB\_BR\_Qual} that are relevant to the reliability of $M_{\rm BH}$ and $L_{\rm bol}$ (see also Table \ref{catalog}).
We find that $L_{\rm 5100}$ and $FWHM_{{\rm H}\beta}$ are securely estimated for 21,888/37,181 ($\sim$58.9\%) objects.
To ensure the accuracy of fitting results, we focus objects with $\chi_{\rm QSFit}^{2} <$ 3.0, {\tt Cont\_5100\_Qual} = 0, and {\tt HB\_BR\_Qual} = 0 for the following analysis (see Section \ref{CF}).

\begin{figure}
    \centering
    \includegraphics[width=0.45\textwidth]{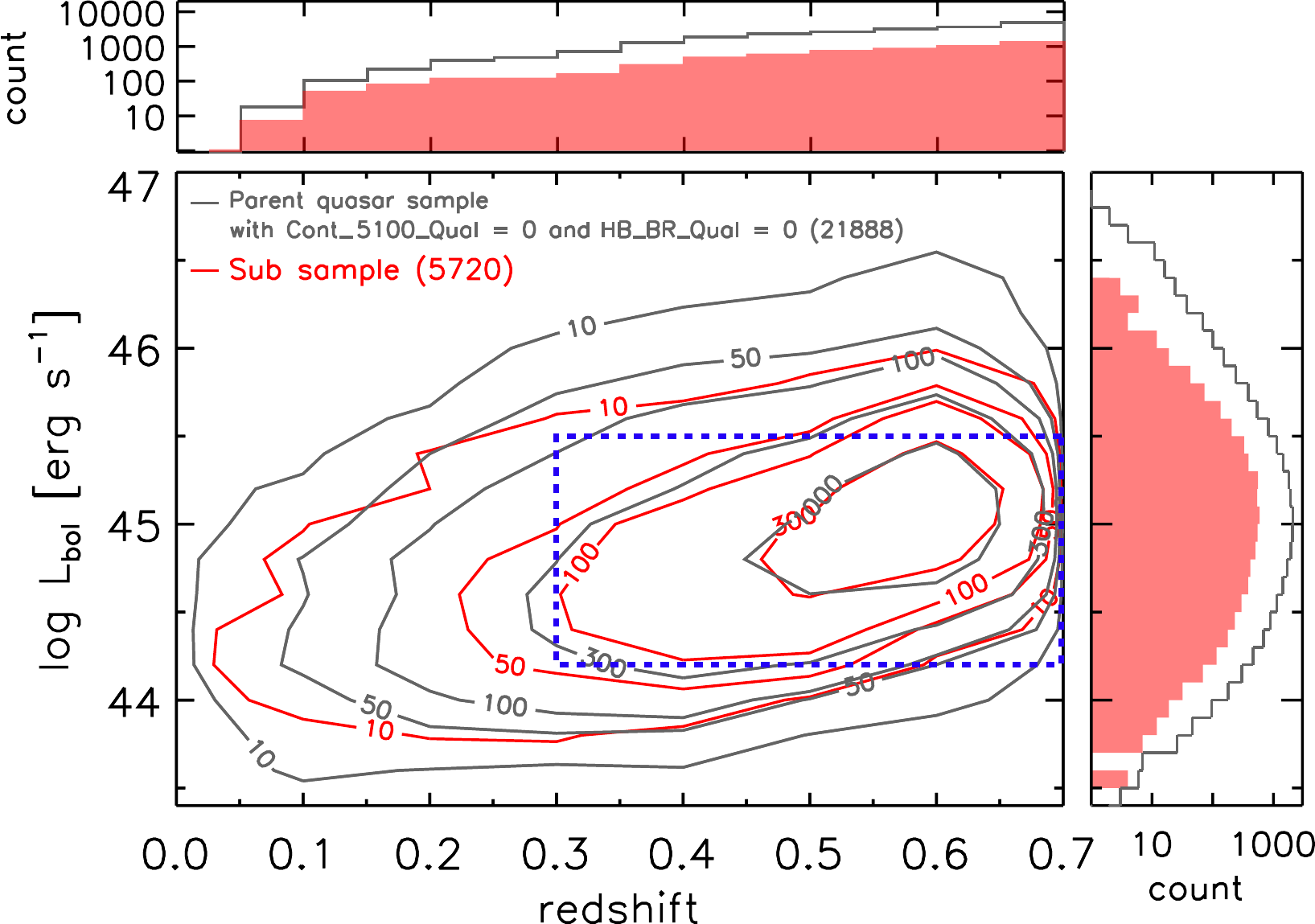}
\caption{The bolometric luminosity as a function of redshift for our sample. The histograms of redshift and $L_{\rm bol}$ are attached on the top and right, respectively. The gray and red contours represent the number density of 21,888 objects with {\tt Cont\_5100\_Qual} = 0 and {\tt HB\_BR\_Qual} = 0, and 5720 objects for correlation analysis, respectively, in each $0.1 \times 0.2$ region on $z-\log L_{\rm bol}$ plane. Objects within the blue dotted square are used to test the influence of the Malmquist bias on correlation analysis (see Section \ref{Malmquist}).}
\label{Dist_phys_z}
\end{figure}

\begin{figure}
    \centering
    \includegraphics[width=0.45\textwidth]{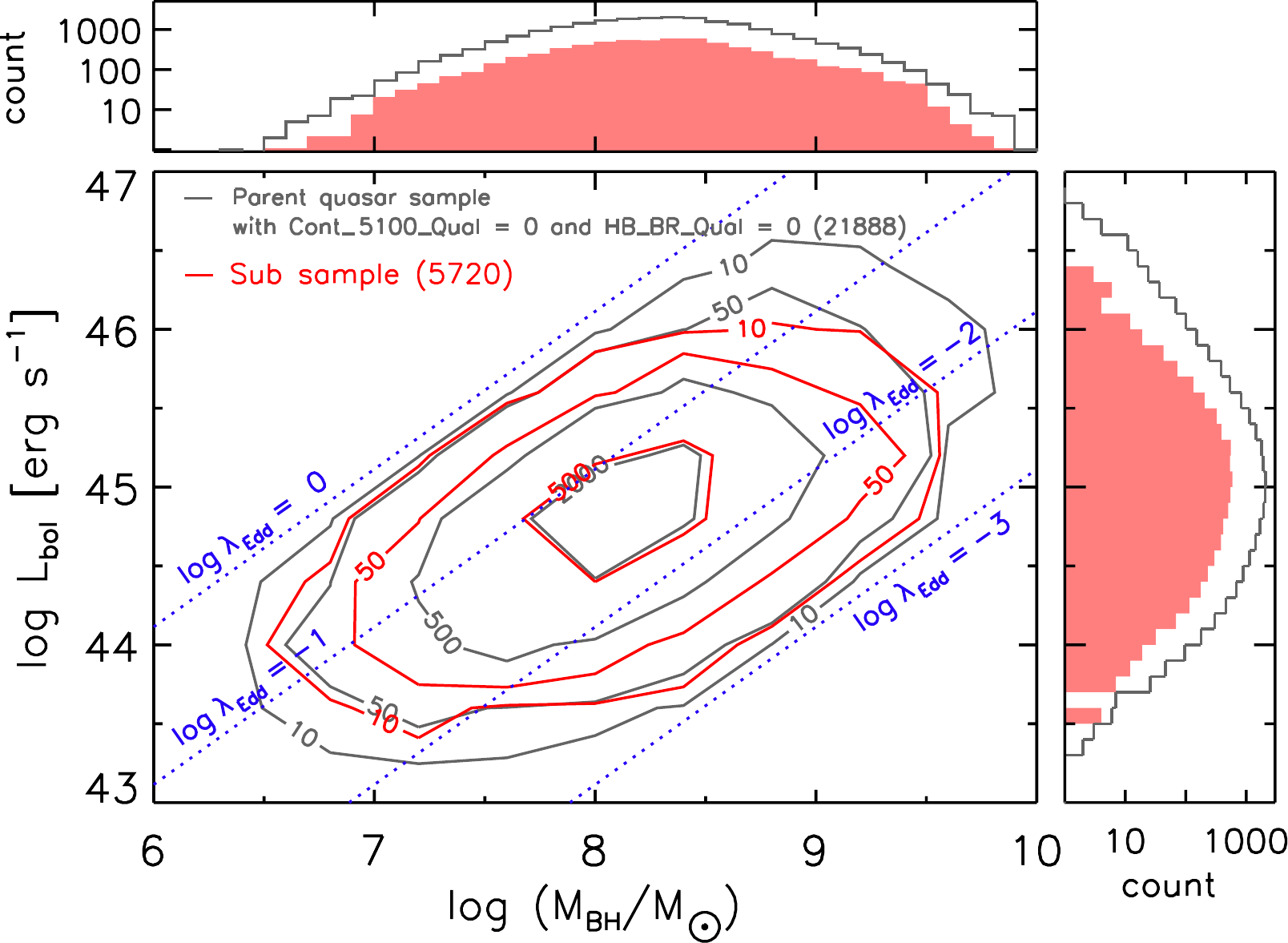}
\caption{The BH mass and bolometric luminosity of our sample. The histograms of $M_{\rm BH}$ and $L_{\rm bol}$ are attached on the top and right, respectively. The gray and red contours represent the number density of 21,888 objects with {\tt Cont\_5100\_Qual} = 0 and {\tt HB\_BR\_Qual} = 0, and 5720 objects for correlation analysis, respectively, in each $0.4 \times 0.4$ region on $\log M_{\rm BH}-\log L_{\rm bol}$ plane. The blue diagonal lines show Eddington ratio of $\log \lambda_{\rm Edd}$ = $-$3, $-$2, $-$1, and 0 from bottom right to top left.}
\label{Dist_phys}
\end{figure}

The distributions of our sample with {\tt Cont\_5100\_Qual} = 0 and {\tt HB\_BR\_Qual} = 0 in $z- L_{\rm bol}$ and $M_{\rm BH}-L_{\rm bol}$ plane are shown in Figures \ref{Dist_phys_z} and \ref{Dist_phys}, respectively.
The mean and standard deviation of $\log\, L_{\rm bol}$~(erg s$^{-1}$), $\log\, (M_{\rm BH}/M_\sun)$, and $\log \,\lambda_{\rm Edd}$ are 45.0 $\pm$ 0.46, 8.29 $\pm$ 0.45, and $-$1.41 $\pm$ 0.38, respectively.

\subsection{Dependences of CF on $L_{\rm bol}$, $M_{\rm BH}$, and $\lambda_{\rm Edd}$}   
\label{CF}

We investigate the dependences of CF on $L_{\rm bol}$, $M_{\rm BH}$, and $\lambda_{\rm Edd}$ of our quasar sample.
As mentioned in Sections \ref{BIC} and \ref{SpF}, quasars with (i) $\chi_{\rm CIGALE}^{2} < 3.0$, (ii) $\Delta$BIC $>$ 6, (iii) $\chi_{\rm QSFit}^{2} <$ 3.0, and (iv) {\tt Cont\_5100\_Qual} = 0, and (v) {\tt HB\_BR\_Qual} = 0 are used for the following correlation analysis, which yields 5720 objects (see also Figure \ref{FC}).
We discuss how the above selection cuts affects the dependences of CF in Section \ref{SB}.
Their distributions of our sub-sample in $z-L_{\rm bol}$ and $M_{\rm BH}-L_{\rm bol}$ plane are also shown in Figures \ref{Dist_phys_z} and \ref{Dist_phys}, respectively.
We find that our quasar sample selected from the DR16Q covers 43.5 $<$ $\log L_{\rm bol}$ (erg s$^{-1}$) $<$ 46.4, 6.57 $<$ $\log (M_{\rm BH}/M_\sun)$ $<$ 9.89, and $-$3.16 $<$ $\log \lambda_{\rm Edd}$ $<$ $-$0.22.
The mean and standard deviation of $\log\, L_{\rm bol}$~(erg s$^{-1}$), $\log\, (M_{\rm BH}/M_\sun)$, and $\log \,\lambda_{\rm Edd}$ are 45.0 $\pm$ 0.39, 8.29 $\pm$ 0.46, and $-$1.41 $\pm$ 0.38, respectively.
The fact that the DR16Q contains fainter quasars than those in previous releases enables us to investigate the CF for less luminous quasars with less massive BH and smaller accretion rate compared to previous studies baaed on the SDSS quasar catalogs \citep[e.g.,][]{Gu,Ma}.

As we cautioned in Section \ref{intro}, the IR-to-bolometric luminosity ratio would not be always a good tracer of  CF.
We hence converted from $L_{\rm IR}^{\rm torus}/L_{\rm bol}$ to CF by using a non-liner relation in \cite{Stalevski} who provides conversion formula assuming {\tt SKIRTOR} with $R_{\rm max}/R_{\rm min}$ = 30 and  	
$\theta$ = 0$\arcdeg$ that are the same in Table \ref{Param} (see also the middle panel of Figure 7 in \citealt{Stalevski});
\begin{eqnarray}
\label{Eq_CF}
 {\rm CF} &=& a_4\left(\frac{L_{\rm IR}^{\rm torus}}{L_{\rm bol}}\right)^4 + a_3\left(\frac{L_{\rm IR}^{\rm torus}}{L_{\rm bol}}\right)^3 \nonumber \\
 &+& a_2\left(\frac{L_{\rm IR}^{\rm torus}}{L_{\rm bol}}\right)^2 + a_1\left(\frac{L_{\rm IR}^{\rm torus}}{L_{\rm bol}}\right) + a_0.
\end{eqnarray}
Since we parameterized $\tau_{\rm 9.7}$ (see Table \ref{Param}), we chose $(a_0, \,a_1, \,a_2, \,a_3, \,a_4)$ = (0.192478, 1.40827, $-$1.48727, 0.875215, $-$0.177798), (0.195615, 1.20218, $-$1.04546, 0.47493, $-$0.0601471), and (0.196387, 1.02696, $-$0.782418, 0.299937, $-$0.0255416) for objects with $\tau_{\rm 9.7}$ = 3, 5, and 9, respectively.
The uncertainty of CF was propagated from a relative error of $L_{\rm IR}^{\rm torus}/L_{\rm bol}$.

\begin{figure*}
    \centering
    \includegraphics[width=0.9\textwidth]{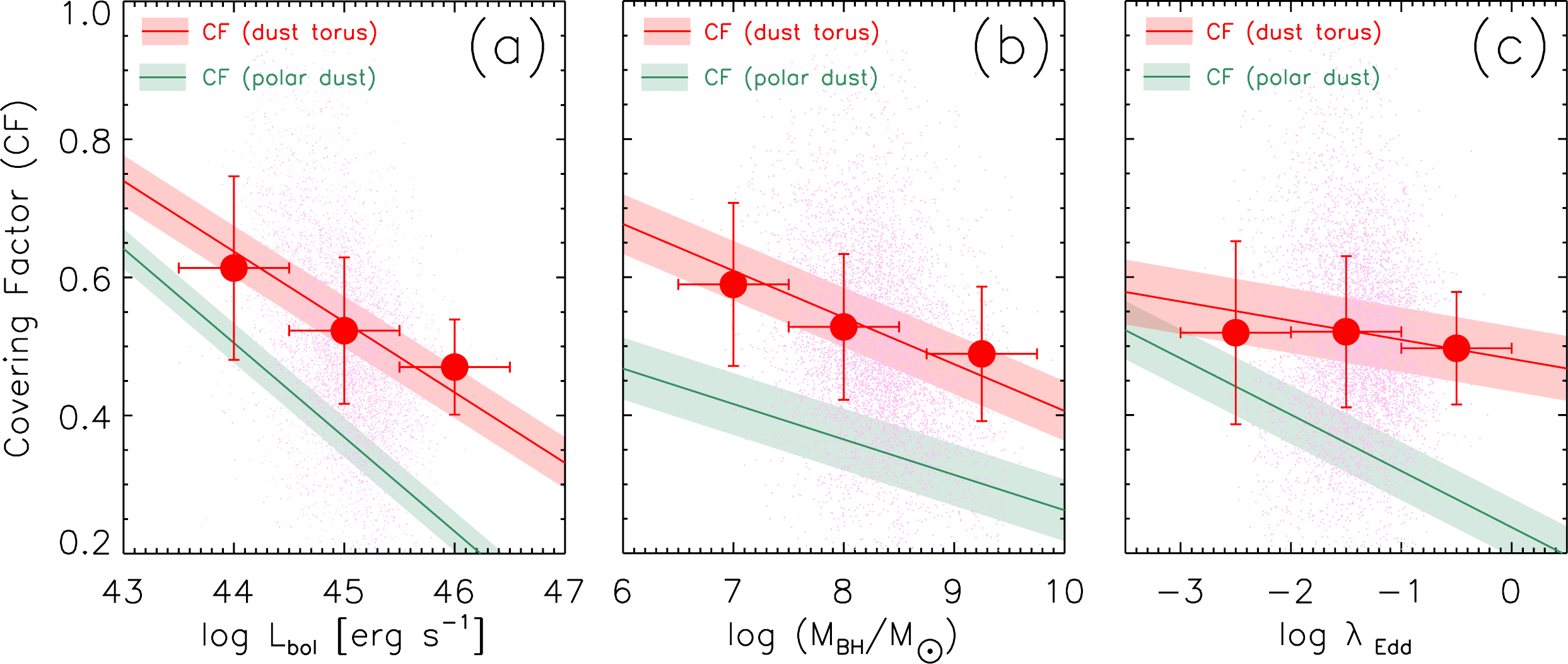}
\caption{The relation between CF and (a) $\log \,L_{\rm bol}$, (b) $\log \,M_{\rm BH}$, and (c) $\log \,\lambda_{\rm Edd}$ for 5720 quasars at $z < 0.7$. Reddish symbols and lines represent CF of dust torus while green lines represent CF of polar dust. Magenta points represent data while red circles represent weighted mean and standard deviation. Red and green lines with shaded region denote the best-fit relations with 1$\sigma$ uncertainty.}
\label{fig_CF}
\end{figure*}

Figure \ref{fig_CF} shows the relation between covering factor of dust torus (CF) and AGN properties, i.e., CF--$L_{\rm bol}$, CF--$M_{\rm BH}$, and CF--$\lambda_{\rm Edd}$ for 5720 quasars.
We confirm an anti-correlation of the above three quantities, which is consistent with recent works \citep{Ezhikode,Ricci,Zhuang}.
We also investigate dependence of CF of polar dust (CF$_{\rm PD}$) on $L_{\rm bol}$, $M_{\rm BH}$, and $\lambda_{\rm Edd}$, which is show in Figure \ref{fig_CF}.
We find that CF$_{\rm PD}$ also clearly depend on AGN properties (see Section \ref{D} for quantitative discussion).

\section{Discussion} 
\label{D}

First, we discuss how the AGN polar dust affects the correlation coefficient of dust CF and AGN properties such as $L_{\rm bol}$, $M_{\rm BH}$, and $\lambda_{\rm Edd}$.
We then argue possible selection effects and uncertainties of CF and correlation analysis through Monte Carlo simulation and mock analysis.
Finally, we discuss the dependence of CF--$\lambda_{\rm Edd}$ correlation on $L_{\rm bol}$, $M_{\rm BH}$, and redshift of our sample.

\subsection{Correlation Analysis}
\label{CA}
To quantify the anti-correction shown in Figure \ref{fig_CF}, we conducted a correlation analysis for the sub-sample of 5720 quasars by using a Bayesian maximum likelihood method provided by \cite{Kelly}, providing a correlation coefficient ($r$) which takes into account uncertainty on both the $x$ and $y$ values \citep[see e.g.,][]{Mateos,Toba_19a}.
The same analysis was done for the same sub-sample of 5720 quasars whose CF was derived by the SED fitting without adding the polar dust component in order to check how the presence or absence of polar dust would affect the correlation strength.
The resultant correlation coefficients are summarized in Table \ref{coeff}.

\begin{deluxetable}{lccc}
\tablenum{3}
\tablecaption{Summary of the correlation coefficient ($r$). \label{coeff}}
\tablewidth{0pt}
\tablehead{
\colhead{}	&	\colhead{$r$ without polar dust}	&	\colhead{$r$ with polar dust}
}
\startdata
CF --$L_{\rm bol}$		&	$-$0.55 $\pm$ 0.01 &	$-$0.73 $\pm$ 0.02\\
CF --$M_{\rm BH}$		&	$-$0.48 $\pm$ 0.01 &	$-$0.58 $\pm$ 0.03\\
CF --$\lambda_{\rm Edd}$&	$-$0.01 $\pm$ 0.02 &	$-$0.21 $\pm$ 0.04\\ 
\hline
CF$_{\rm PD}$--$L_{\rm bol}$		&	---		&	$-$0.88 $\pm$ 0.01\\
CF$_{\rm PD}$--$M_{\rm BH}$			&	---		&	$-$0.46 $\pm$ 0.03\\
CF$_{\rm PD}$--$\lambda_{\rm Edd}$	&	---		&	$-$0.58 $\pm$ 0.02\\ 
\enddata
\tablecomments{CF and CF$_{\rm PD}$ denote covering factor of dust torus and polar dust, respectively.}
\end{deluxetable}
\begin{figure}
    \centering
    \includegraphics[width=0.4\textwidth]{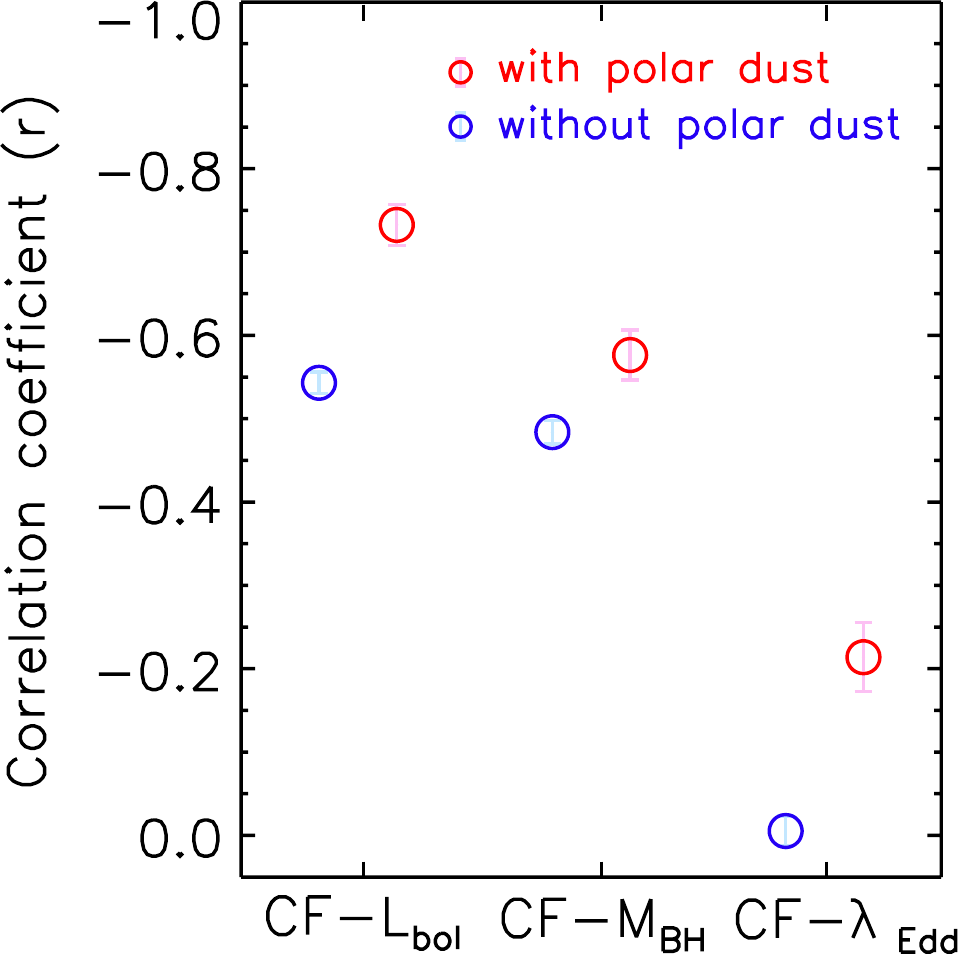}
\caption{Correlation coefficient ($r$) of CF--$L_{\rm bol}$, CF-- $M_{\rm BH}$, and CF--$\lambda_{\rm Edd}$. $r$ with and without adding AGN polar dust component to the SED fitting are plotted as red and blue color, respectively.}
\label{CC}
\end{figure}

Figure \ref{CC} shows correlation coefficient ($r$) of CF--$L_{\rm bol}$, CF-- $M_{\rm BH}$, and CF--$\lambda_{\rm Edd}$ with and without adding AGN polar dust component to the SED fitting.
We find that the $|r|$ values of objects when considering polar dust emission are larger than those without considering the polar dust.
In particular, the influence of the polar dust component on the correlation strength may be significant for CF--$\lambda_{\rm Edd}$ rather than CF--$L_{\rm bol}$ and CF--$M_{\rm BH}$. 
This indicates that polar dust wind, probably driven by radiation pressure from the AGN may be crucial for regulating  obscuring dusty structure surrounding SMBHs.
Table \ref{coeff} also provides insight into property of polar dust; CF$_{\rm PD}$ seems to strongly depend on $L_{\rm bol}$ and $\lambda_{\rm Edd}$ rather than $M_{\rm BH}$, which is consistent with what we discussed the above.
Strong radiation pressure from luminous quasars with high $\lambda_{\rm Edd}$ may blowout dust in polar direction, which would be associate with an AGN-driven outflow \citep[e.g.,][]{Schartmann14}.
This suggests that CF$_{\rm PD}$ is regulated by $L_{\rm bol}$ and possibly $\lambda_{\rm Edd}$.

\subsection{Selection Bias}
\label{SB}

\subsubsection{Parent SDSS quasar sample}
\label{parent}

We used 37,181 quasars at $z < 0.7$ drawn from the SDSS DR16Q as a parent sample (see Figure \ref{FC}).
Since the completeness and contamination rate of the parent sample would directly affect our correlation analysis, we estimate them following \cite{Lyke}.
We utilized 323 objects with {\tt RANDOM\_SELECT} = 1 that are randomly selected from the DR16Q.
This subsample was visually inspected to check whether the pipeline correctly classified the spectrum.
If an object is visually confirmed to be a quasar, the object has {\tt IS\_QSO\_10K} = 1.
The confidence rating for a visually identified redshift is stored in {\tt Z\_CONF\_10K}.
We consider objects with {\tt Z\_CONF\_10K} $\geq$ 2 as those with correct redshift \citep[see Section 3.3 in][in detail]{Lyke}.

The completeness and contamination rate of the parent sample estimated based on Equations (1) and (2) in \cite{Lyke} is 99.7\% and 0.3\%, respectively.
This means that the parent quasar sample in this work has quite high completeness and purity, which may not affect our conclusion in this work.

\subsubsection{Malmquist bias}
\label{Malmquist}

Because the target selection for quasars was proceeded from flux-limited samples \citep[e.g.,][]{Myers}, our correlation analysis would be affected by Malmquist bias (see also Figure \ref{Dist_phys_z}).
In order to see how the Malmquist bias would affect the correlation coefficients, we created a sub-sample that is expected not to be affected by the Malmquist bias in which 4785 objects with $44.2 < \log\, L_{\rm bol} < 45.5$ and $z > 0.3$ were extracted (see a blue dotted square in Figure \ref{Dist_phys_z}).
We then conducted the correlation analysis for the sub-sample in the same manner as what is described in Section \ref{CA}.
We find that the mean and standard deviations of $r$ for CF--$L_{\rm bol}$, CF--$M_{\rm BH}$, and CF--$\lambda_{\rm Edd}$ are $-$0.92 $\pm$ 0.05, $-$0.69 $\pm$ 0.10, and $-$0.30 $\pm$ 0.15, respectively.
The absolute values of $r$ tend to be higher than what is reported in Table \ref{coeff} although there are large standard deviations particularly for $r ({\rm CF}-M_{\rm BH}$) and $r ({\rm CF}-\lambda_{\rm Edd}$). 
This result would indicate that Malmquist bias makes the correlation strength weaker.

\subsubsection{Sub-sample of quasars}
\label{sub_q}
\begin{figure}
    \centering
    \includegraphics[width=0.4\textwidth]{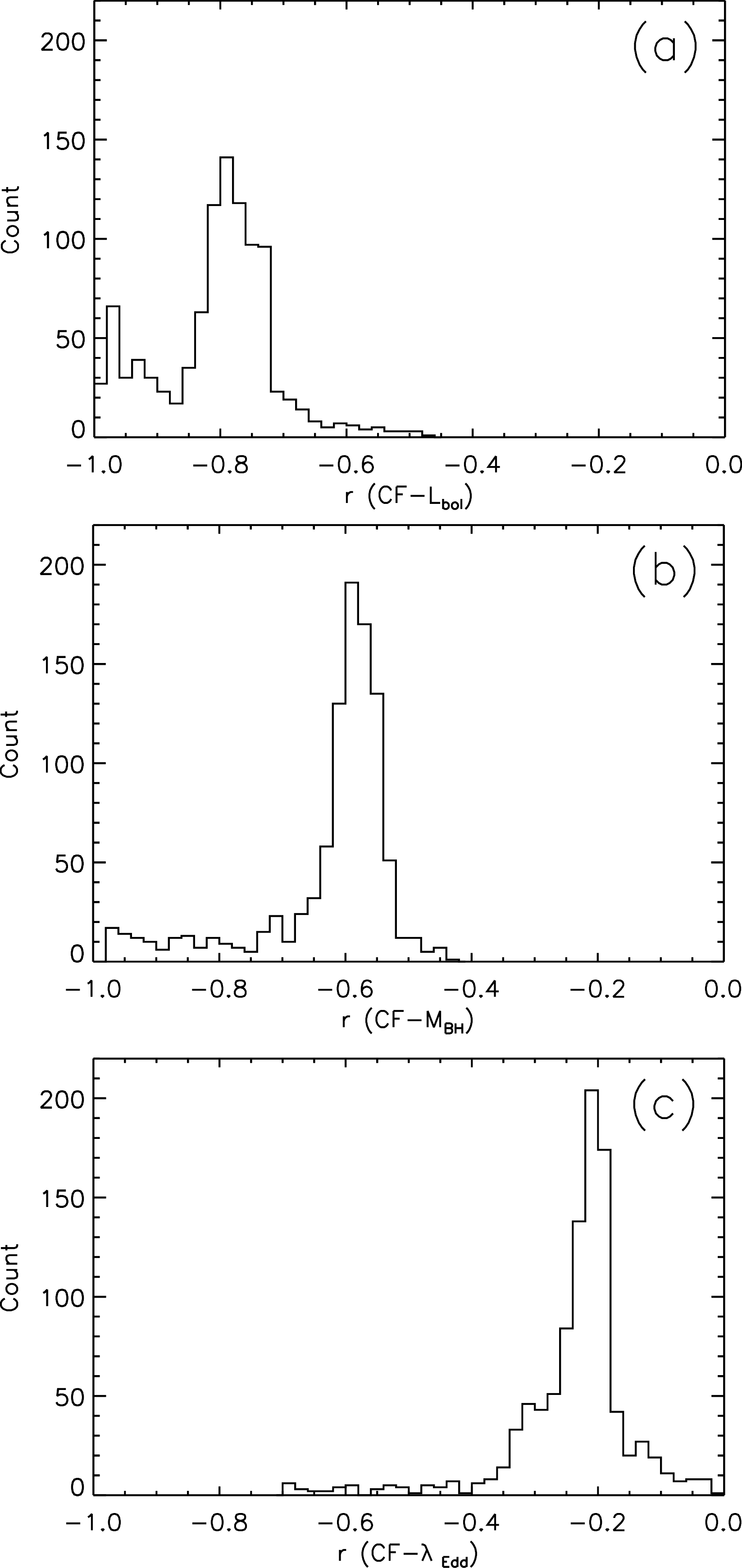}
\caption{Simulated distributions of correlation coefficients for (a) CF--$L_{\rm bol}$, (b) CF--$M_{\rm BH}$, and (c) CF--$\lambda_{\rm Edd}$ in which a set of threshold value for sample selection (reduced $\chi^2_{\rm CIGALE}$, reduced $\chi^2_{\rm QSFit}$, and $\Delta$BIC) was randomly chosen (see Section \ref{sub_q}).}
\label{Monte}
\end{figure}
As described in Sections \ref{BIC} and \ref{SpF}, we created a sub-sample for the correlation analysis by considering the reduced $\chi^2_{\rm CIGALE}$, $\chi^2_{\rm QSFit}$, $\Delta$BIC, {\tt Cont\_5100\_Qual}, and {\tt HB\_BR\_Qual} to understand how AGN polar dust affects dust CF.
In particular, the threshold of $\Delta$BIC contributed to a decrease in the sample size of quasars from 37,181 to 5720, although the number of the sample is still enough for statistical investigation.
We discuss how the selection criteria for the correlation analysis would be biased toward the resultant correlation coefficient.
To address this issue, we performed a Monte Carlo simulation in which we randomly chose a set of threshold value (reduced $\chi^2_{\rm CIGALE}$, reduced $\chi^2_{\rm QSFit}$, and $\Delta$BIC), and conducted the correlation analysis.
We iterated the above process 1000 times where calculation was executed only if the sample size exceeds 1000 in order to insure a reliability of resultant correlation coefficient.

Figure \ref{Monte} shows the distribution of $r$. 
The mean and standard deviations of $r$ for CF--$L_{\rm bol}$, CF--$M_{\rm BH}$, and CF--$\lambda_{\rm Edd}$ are $-$0.80 $\pm$ 0.09, $-$0.62 $\pm$ 0.10 and $-$0.23 $\pm$ 0.10, respectively, which are roughly consistent with what we obtained in this work (see Table \ref{coeff}).
This suggests that selection bias caused by adopting threshold values in this work does not significantly affect the result of the correlation analysis.

\subsection{Influence of a limited number of stellar templates on the correlation coefficients}
\label{age}

For the SED and spectral fitting in a consistent manner, we fixed the age of the main stellar population to be 5.0 Gyr for the SED fitting while we used the template of an elliptical galaxy with a stellar age of 5.0 Gyr for the spectral fitting (see Sections \ref{cigale} and \ref{qsfit}).
Although this assumption (i.e., low-z quasars are hosted by elliptical galaxies) is expected to apply to the majority of our quasar sample \citep[e.g.,][]{Bahcall,Dunlop,Kauffmann,Floyd}, a fraction of quasars could not be the case \citep[e.g.,][]{McLure,Schawinski,Ishino}.
How do these possible variations of quasar host properties would affect resulting correlation coefficients?

To address this issue, we modified (i) the parameter ranges in the age of the main stellar population used for the SED fitting and (ii) AGN host templates for the spectral fitting.
For the SED fitting, we allowed the age of the main stellar population to vary from 1.0 to 11.0 Gyr at intervals of 2.0 Gyr.
For the spectral fitting, we utilized host galaxy templates of a simulated 2.0 Gyr old elliptical galaxy, S0-galaxy, and three types of spiral galaxies (i.e., Sa, Sb, and Sc) \citep{Polletta}, in addition to the 5.0 Gyr old  elliptical galaxy template we originally used.
We randomly chose a host galaxy template from the above options.
We then performed the SED fitting and spectral fitting in the same manner as what is described in Sections \ref{cigale} and \ref{qsfit}.
We iterated this process 1000 times and estimate the distribution of $r$.

\begin{figure}
    \centering
    \includegraphics[width=0.4\textwidth]{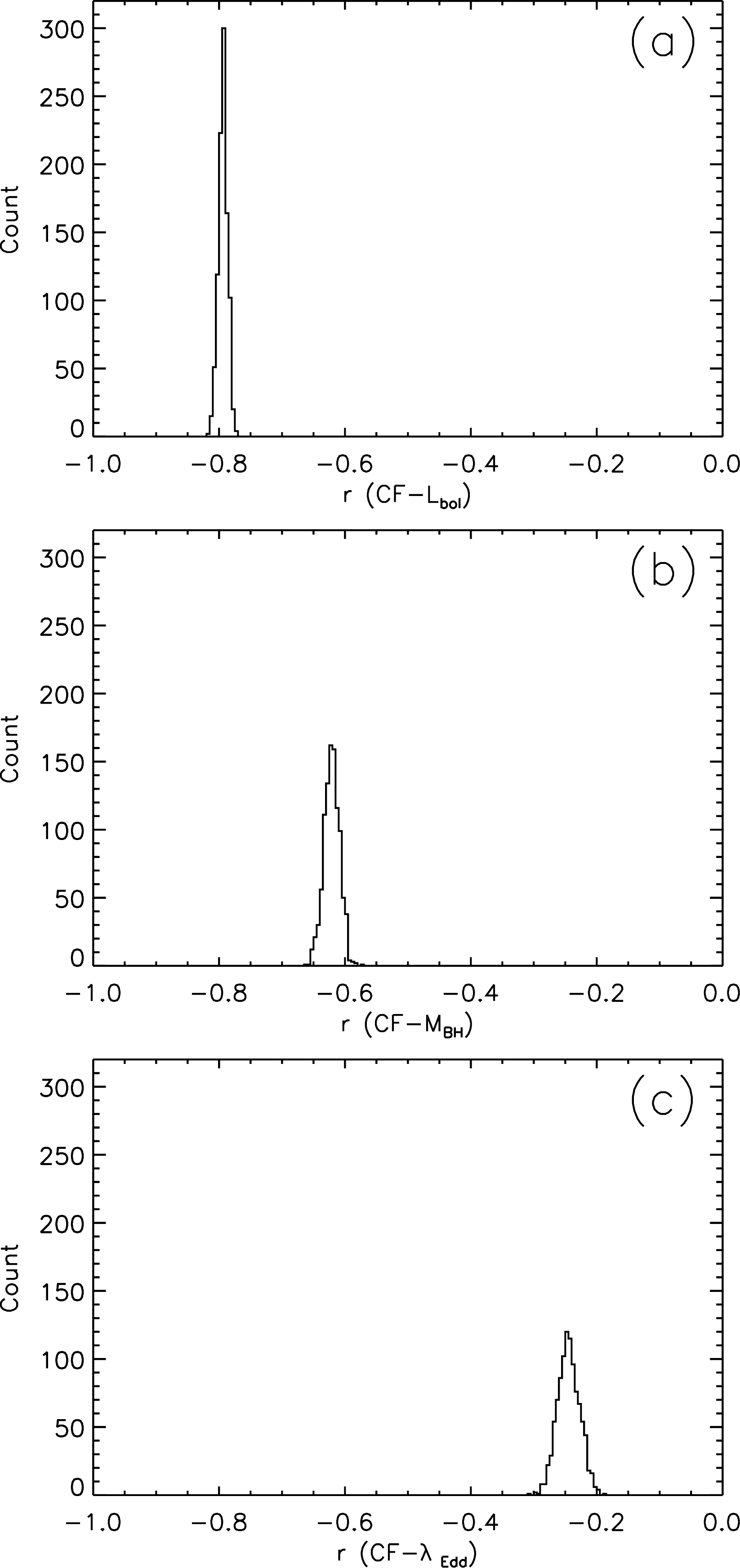}
\caption{Simulated distributions of correlation coefficients for (a) CF--$L_{\rm bol}$, (b) CF--$M_{\rm BH}$, and (c) CF--$\lambda_{\rm Edd}$ in which stellar template for spectral fitting was randomly chosen (see Section \ref{age}).}
\label{Monte_temp}
\end{figure}
The resulting distributions of $r$ are shown in Figure \ref{Monte_temp}.
We find that the mean and standard deviations of $r$ for CF--$L_{\rm bol}$, CF--$M_{\rm BH}$, and CF--$\lambda_{\rm Edd}$ are $-$0.79 $\pm$ 0.00, $-$0.62 $\pm$ 0.01, and $-$0.24 $\pm$ 0.01, respectively, which are in good agreement   with what we obtained in this work (see Table \ref{coeff}).
Therefore, we conclude that the influence of a limited number of stellar templates on the correlation coefficients is likely to be small.

\subsection{Influence of CF--$L_{\rm bol}$ correlation on CF--$\lambda_{\rm Edd}$ correlation coefficient}
\label{inf_Lbol}

We reported in Section \ref{CA} that CF depends on $\lambda_{\rm Edd}$ by taking into account polar dust emission.
It should be noted that $\lambda_{\rm Edd}$ is directly related to $L_{\rm bol}$ that strongly correlates with CF, which would induce an artificial correlation of CF--$\lambda_{\rm Edd}$.

To disentangle the dependence and see how $r$ (CF--$\lambda_{\rm Edd}$) would be affected by strong correlation of CF--$L_{\rm bol}$, we employed a diagnostic method presented in \cite{Hasinger} who investigated redshift dependence on CF based on the relation between the CF and X-ray luminosity \citep[see also][]{Toba_14}.
First, we estimated the average value of the slope of the relation between the CF and $L_{\rm bol}$ in the range of $-2.0 < \log\,\lambda_{\rm Edd} < 0.0 $.
We then estimated the normalization value at a luminosity of $\log\,L_{\rm bol}$ = 45.0 erg s$^{-1}$, in the middle of the observed range, as a function of the $\lambda_{\rm Edd}$ by keeping the slope fixed to the average value.
The $\lambda_{\rm Edd}$ bins were $-2.0 < \log\, \lambda_{\rm Edd} < -1.5$, $-1.5 < \log\, \lambda_{\rm Edd} < -1.0$, $-1.0 < \log\,\lambda_{\rm Edd} < -0,5$, and $-0.5 < \log\,\lambda_{\rm Edd} < 0.0$ in the range of $\log\,L_{\rm bol}$ = 44--46 erg s$^{-1}$.
We find that the resulting correlation coefficient is $r$ (CF--$\lambda_{\rm Edd}$) $\sim$ 0.25, which is consistent with what we obtained in Section \ref{CA}.
Therefore, we conclude that a strong CF--$L_{\rm bol}$ correlation may not significantly affect the correlation coefficient of CF--$\lambda_{\rm Edd}$.

\subsection{Mock Analysis}
\label{mock}
Finally, we check whether or not the derived CF can actually be estimated reliably, given the limited number of photometry and its uncertainty.
We execute a mock analysis that is a procedure provided by {\tt X-CIGALE} \citep[see e.g.,][for more detail]{Buat_12,Buat_14,Ciesla_15,LoFaro,Boquien,Toba_19b,Toba_20c} for 5720 quasars. 
This analysis is done by a mock catalog in which a value is taken from a Gaussian distribution with the same standard deviation as the observation is added to each photometry originally used, which would also enable to test how the difference in photometry influences CF (see Section \ref{cigale}).

\begin{figure}
    \centering
    \includegraphics[width=0.4\textwidth]{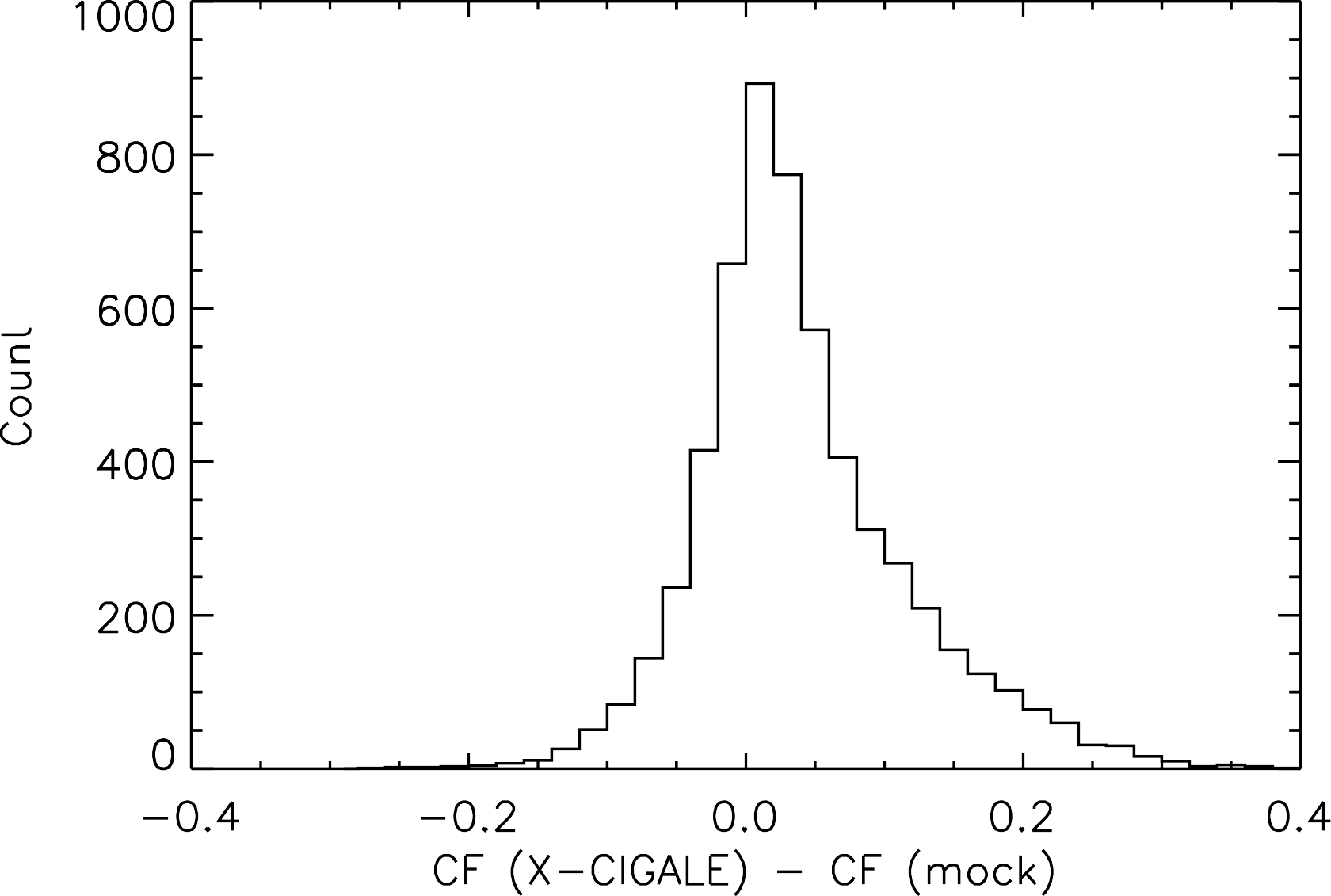}
\caption{Histogram of difference in CFs that are derived in this work and from mock analysis for 5720 quasars.}
\label{fig_mock}
\end{figure}

Figure \ref{fig_mock} shows the histogram of differences in CFs that are derived from {\tt X-CIGALE} in this work and from the mock catalog.
The mean and standard deviations of $\Delta$CF = 0.04 $\pm$ 0.07, suggesting that CF of majority of our quasar sample is insensitive to the limited number of photometric points and difference in photometry.

\subsection{Influence of AGN properties on CF--$\lambda_{\rm Edd}$ Correlation}
\label{Comp}

In section \ref{CA}, we report that adding AGN polar dust emission to SED fitting makes anti-correlations of CF--$L_{\rm bol}$, CF--$M_{\rm BH}$, and CF--$\lambda_{\rm Edd}$ more stronger than those without adding it.
On the other hand, $L_{\rm bol}$, $M_{\rm BH}$, $\lambda_{\rm Edd}$, and also redshift should be correlated with each other as mentioned in Sections \ref{Malmquist} and \ref{inf_Lbol} (see also Figures \ref{Dist_phys_z} and \ref{Dist_phys}).
With all the caveats discussed in Sections \ref{SB} -- \ref{mock} in mind, we discuss how CF--$\lambda_{\rm Edd}$ correlation depends on AGN properties (i.e., $L_{\rm bol}$ and $M_{\rm BH}$) and redshift.
The resultant correlation coefficients are summarized in Table \ref{coeff2}.

\begin{deluxetable}{llc}
\tablenum{4}
\tablecaption{Dependence of correlation coefficient of CF--$\lambda_{\rm Edd}$ on redshift, BH mass, and bolometric luminosity.
\label{coeff2}}
\tablewidth{0pt}
\tablehead{
\colhead{}	&	\colhead{Parameter Range}	&	\colhead{$r$ (CF--$\lambda_{\rm Edd}$)}
}
\startdata
 					&	$z < 0.1$ 			&	$-0.33 \pm 0.45$ \\
 Redshift			&	$0.1 < z < 0.2$		&	$-0.25 \pm 0.13$ \\
					&	$z > 0.2$			&	$-0.22 \pm 0.05$ \\	
\hline		
 					&	$\log \,(M_{\rm BH}/M_\sun) < 8$			&	$-0.55 \pm 0.07$\\
 BH mass			&	$8 < \log \,(M_{\rm BH}/M_\sun) < 9$		&	$-0.49 \pm 0.04$\\
					&	$\log \,(M_{\rm BH}/M_\sun) > 10$			&	$-0.45 \pm 0.21$\\	
\hline		
 								&	$\log L_{\rm bol} > 44.0$ 		&	$-0.12 \pm 0.10$\\
 $L_{\rm bol}$ (erg s$^{-1}$)	&	$44 < \log L_{\rm bol} < 45.0$ 	&	$-0.07 \pm 0.12$\\
								&	$\log L_{\rm bol} > 45.0$ 		&	 $0.05 \pm 0.06$\\							
\enddata
\tablecomments{Polar dust component is taken into account to estimate $r$ (CF--$\lambda_{\rm Edd}$).}
\end{deluxetable}

\subsubsection{Redshift Dependence}
\label{SS_z}

Although we confirm that anti-correlations of CF and $L_{\rm bol}$, $M_{\rm BH}$, and $\lambda_{\rm Edd}$, the correlation strength differs among them. 
We find that CF--$L_{\rm bol}$ correlation is stronger than other two correlations as shown in Table \ref{coeff} and Figure \ref{CC}, which seems inconsistent with X-ray based work \citep{Ricci} who reported that CF--$\lambda_{\rm Edd}$ correlation is stronger than CF--$L_{\rm bol}$ correlation.
What causes this discrepancy?

There are some possibilities such as the difference in (i) redshift, (ii) sample selection (i.e., our sample is limited to type 1 quasars while sample in \citealt{Ricci} is not the case), and (iii) definition of CF (i.e., $L_{\rm IR}^{\rm torus}/L_{\rm bol}$ is employed in this work while fraction of obscured sources is employed in \citealt{Ricci}).
It is hard to quantify, however, the influence of the difference in the sample selection and definition of CF on the correlation strength and is beyond the scope of this work.
We thus argue that a possibility that the difference in redshift would cause the discrepancy. 
\cite{Ricci} targeted nearby AGNs with a median redshift of $z\sim0.037$ whereas we focus on AGNs with much higher redshift up to $z = 0.7$.
\begin{figure}
    \centering
    \includegraphics[width=0.4\textwidth]{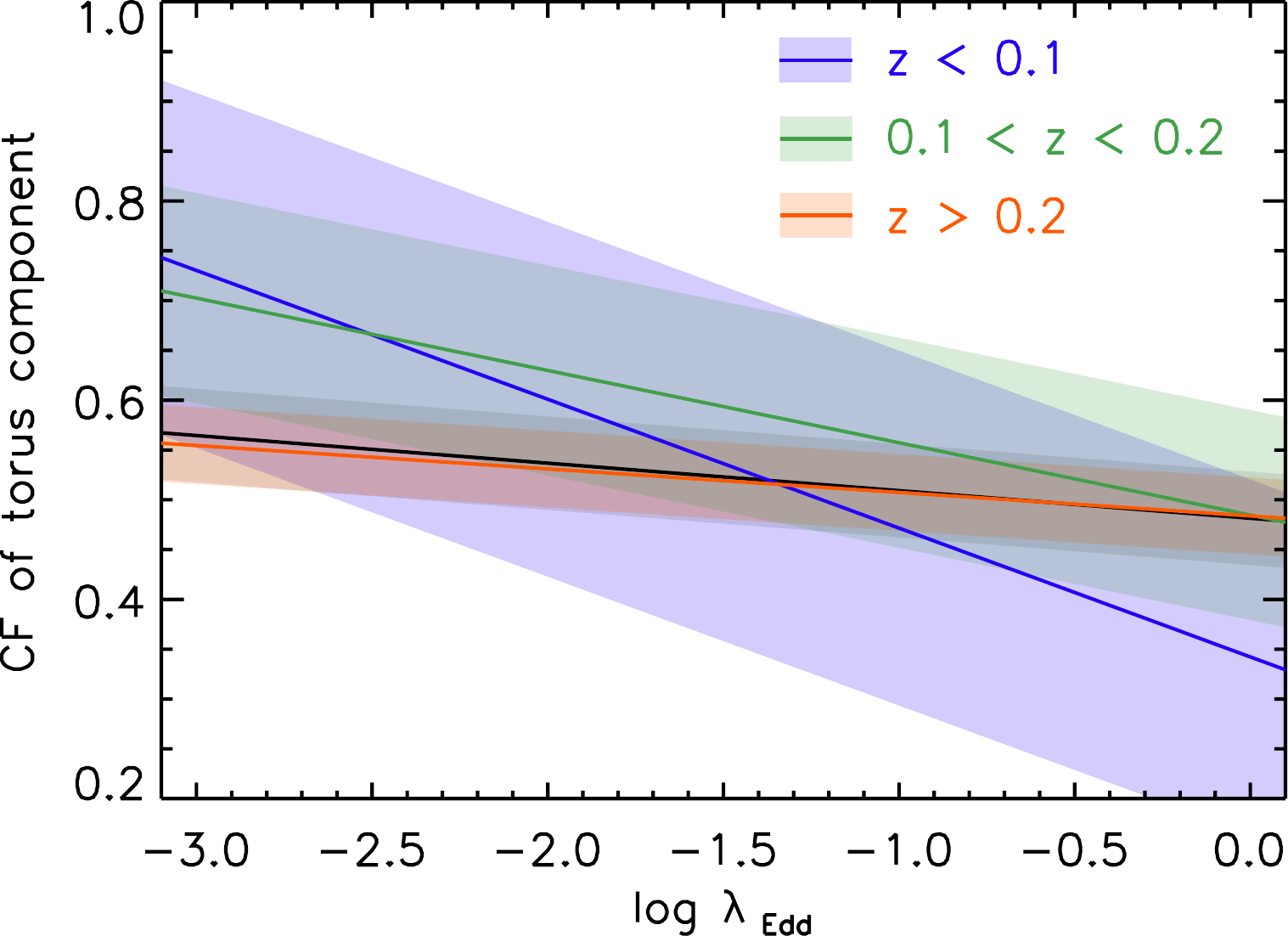}
\caption{The CF as a function of $\lambda_{\rm Edd}$. The blue, green, orange, and gray shaded region represent quasars with $z < 0.1$, $0.1 < z < 0.2$, and $z > 0.2$, and all sample, respectively.}
\label{Comp_z}
\end{figure}

Figure \ref{Comp_z} shows redshift dependence of CF--$\lambda_{\rm Edd}$ relation.
We find that CF of AGNs with $z < 0.1$ (that is similar to that of sample in \cite{Ricci}) tends to strongly depend on $\lambda_{\rm Edd}$, with a correlation coefficient of $r = -0.33 \pm 0.45$ while CF--$\lambda_{\rm Edd}$ for AGNs at $z > 0.2$ shows weaker correlations than what at $z < 0.1$.
This indicates that the aforementioned discrepancy can be explained by difference in redshift of sample between \cite{Ricci} and this work.
We also remind that our correlation analysis would be affected by the Malmquist bias, which could make the correlation strength weaker, as discussed in Section \ref{Malmquist}.

This result also might suggest that CF decreases with increasing redshift at least $z < 0.7$, trend of which is consistent with what is reported in \cite{Toba_14} although there is large uncertainty.
On the other hand, the fraction of obscured X-ray AGNs increases with redshift for the same luminosity \citep[e.g.,][]{Ueda14}.
This discrepancy might be caused by difference in sample selection and definition of CF.

\subsubsection{BH Mass Dependence}
Recently, \cite{Kawakatu} constructed a model of a nuclear starburst disk supported by the turbulent pressure from SNe II and investigated how CF depends on AGN properties by taking account anisotropic radiation pressure from AGNs.
They found that CF strongly depend on $\lambda_{\rm Edd}$ for AGNs with $M_{\rm BH} < 10^8\,M_\sun$ while the dependence of CF on $\lambda_{\rm Edd}$ is much weaker for AGNs with $M_{\rm BH} > 10^{9-10}\,M_\sun$.
Although they did not incorporate the AGN polar dust outflow, it is worth checking the dependence of CF--$\lambda_{\rm Edd}$ correlation on $M_{\rm BH}$.
\begin{figure}
    \centering
    \includegraphics[width=0.4\textwidth]{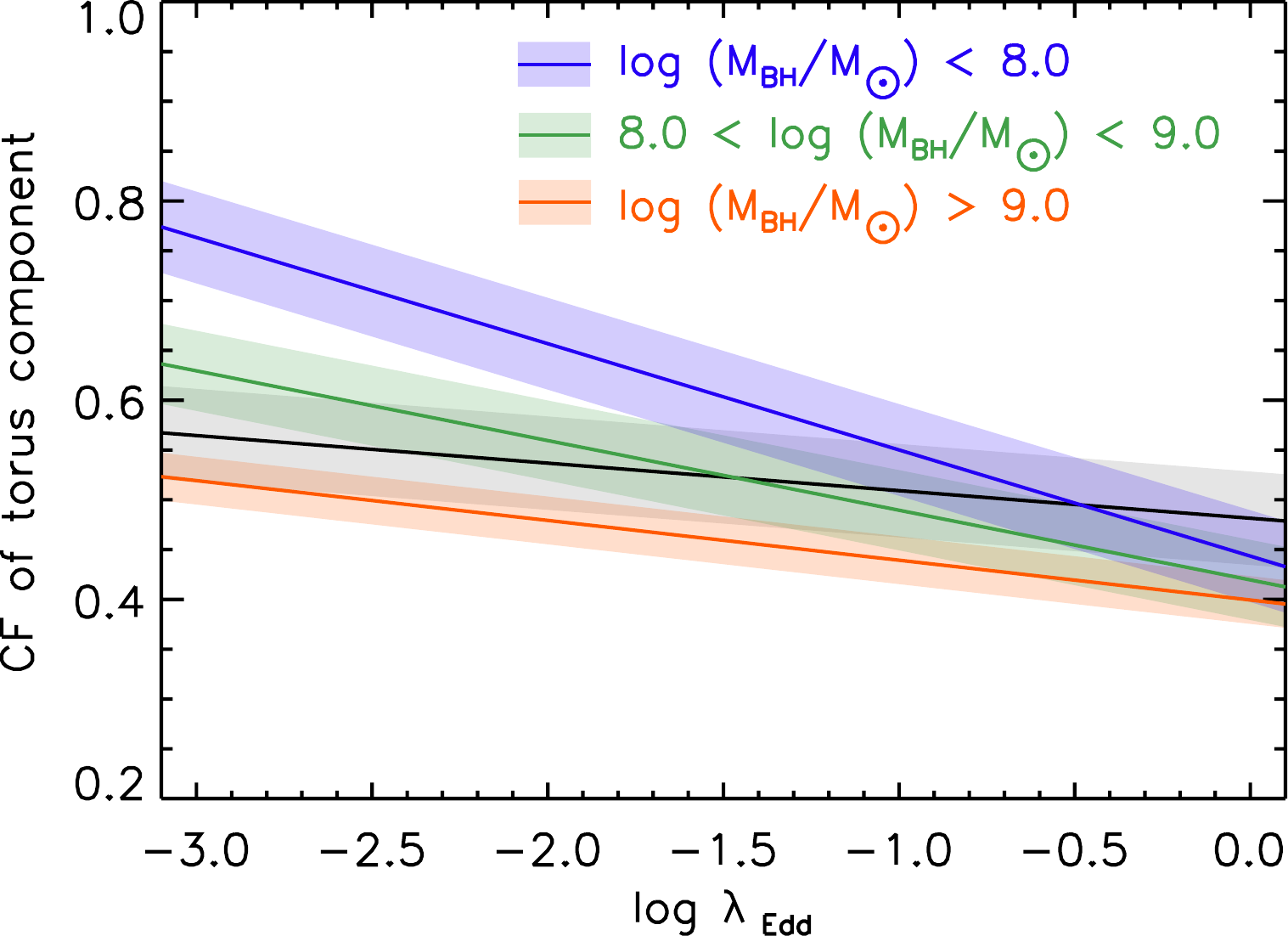}
\caption{The CF as a function of $\lambda_{\rm Edd}$. The blue, green, orange, and gray shaded region represent quasars with $\log\,(M_{\rm BH}/M_\sun) < 8$, $8 < \log\,(M_{\rm BH}/M_\sun) < 9$, $\log\,(M_{\rm BH}/M_\sun) > 9$, and all sample, respectively.}
\label{Comp_MBH}
\end{figure}
Figure \ref{Comp_MBH} shows how the difference in $M_{\rm BH}$ would affect on the correlation strength of CF--$\lambda_{\rm Edd}$.
The resultant $r$ of CF--$\lambda_{\rm Edd}$ for quasars with $\log\,(M_{\rm BH}/M_\sun) < 8$, $8 < \log\,(M_{\rm BH}/M_\sun) < 9$, and $\log\,(M_{\rm BH}/M_\sun) > 9$ are $-$0.55 $\pm$ 0.07, $-$0.49 $\pm$ 0.04, and $-$0.45 $\pm$ 0.21, respectively.
This result indicates that CF of quasar with $< 10^8$ $M_\sun$ strongly depends on $\lambda_{\rm Edd}$.
The overall statistical trend is in good agreement with what reported in \cite{Kawakatu}.

\subsubsection{Bolometric Luminosity Dependence}

\begin{figure}
    \centering
    \includegraphics[width=0.4\textwidth]{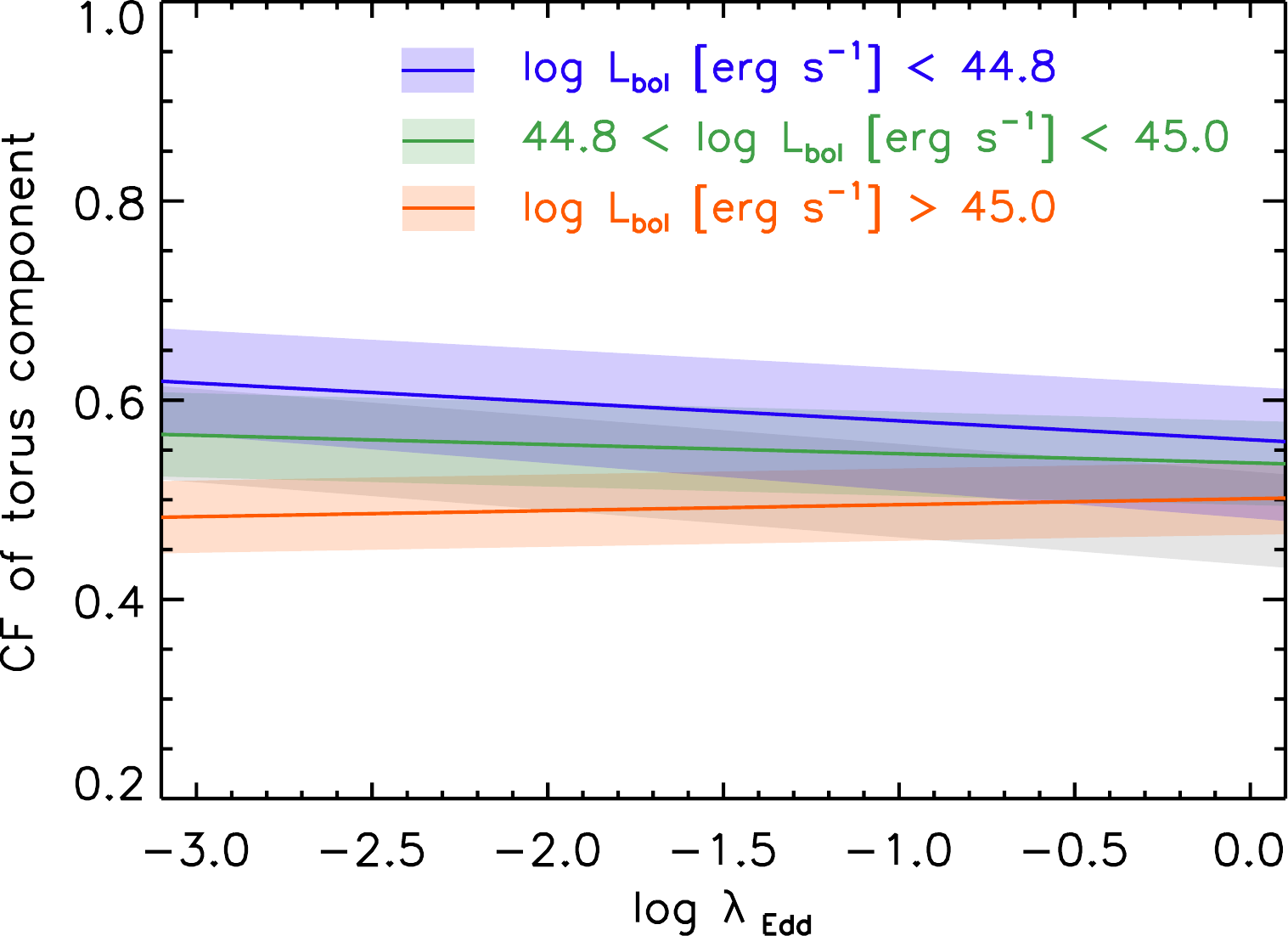}
\caption{The CF as a function of $\lambda_{\rm Edd}$. The blue, green, orange, and gray shaded region represent quasars with $\log\,L_{\rm bol} < 44.8$, $44.8 < \log\, L_{\rm bol} < 45.0$, $\log\, L_{\rm bol} > 45.0$, and all sample, respectively.}
\label{Comp_Lbol}
\end{figure}

Figure \ref{Comp_Lbol} shows dependence of $L_{\rm bol}$ on CF--$\lambda_{\rm Edd}$.
We find that anti-correlation of CF--$\lambda_{\rm Edd}$ may disappear for AGNs with $\log \,L_{\rm bol} > 45.0$.
In this $L_{\rm bol}$ range, the correlation coefficients are even positive with large uncertainty (see Table \ref{coeff2}).
\cite{Zhuang} reported that CF of luminous quasars with $L_{\rm bol} \gtrsim 10^{45.0}$ erg s$^{-1}$ increases with increasing $\lambda_{\rm Edd}$, although this positive correlation would appear when $\log\, \lambda_{\rm Edd}$ ranges from $-$0.25 and 0.5.
This result could indicate that overall trend of CF--$\lambda_{\rm Edd}$ reported in Section \ref{CA} is determined by AGNs with $44.0 < L_{\rm bol} < 45.0$.

\section{Summary} 
\label{Sum}

In this paper, we revisit the relationship between the CF and AGN properties ($L_{\rm bol}$, $M_{\rm BH}$, and $\lambda_{\rm Edd}$) for type 1 quasars selected from the SDSS DR16 quasar catalog.
Thanks to the newly available DR16Q, we can investigate the dependence of CF for quasars with a wide range in $L_{\rm bol}$, $M_{\rm BH}$, and $\lambda_{\rm Edd}$.
We narrowed down the DR16Q catalog to 37,181 quasars with $z < 0.7$, and performed the SED fitting with {\tt X-CIGALE} to at most 13 optical--MIR photometry and spectral fitting with {\tt QSFit} to the SDSS spectra.
In particular, we took into account the contribution of AGN polar dust emission to IR luminosity, which could affect  the measurement of CF.
For 5720 quasars whose physical quantities estimated by {\tt X-CIGALE} and {\tt QSFit} are reliable, we conducted a correlation analysis to see how the AGN polar dust would affect the correlation coefficient.  
We find that taking intro account the contribution of AGN polar dust to IR emission provides stronger anti-correlations of CF--$L_{\rm bol}$, CF--$M_{\rm BH}$, and CF--$\lambda_{\rm Edd}$ than not considering the polar dust.
This result indicates that AGN polar dust wind is a key ingredient to regulate the obscuring structure of AGNs.

\acknowledgments
The authors gratefully acknowledge the anonymous referee for a careful reading of the manuscript and very helpful comments.
We also appreciate Dr. Guang Yang for helping the installation and modification of {\tt X-CIGALE}.

Funding for the Sloan Digital Sky Survey IV has been provided by the Alfred P. Sloan Foundation, the U.S. Department of Energy Office of Science, and the Participating Institutions. SDSS-IV acknowledges support and resources from the Center for High-Performance Computing at the University of Utah. The SDSS web site is www.sdss.org.

SDSS-IV is managed by the Astrophysical Research Consortium for the Participating Institutions of the SDSS Collaboration including the Brazilian Participation Group, the Carnegie Institution for Science, Carnegie Mellon University, the Chilean Participation Group, the French Participation Group, Harvard-Smithsonian Center for Astrophysics, Instituto de Astrof\'isica de Canarias, The Johns Hopkins University, Kavli Institute for the Physics and Mathematics of the Universe (IPMU) / University of Tokyo, the Korean Participation Group, Lawrence Berkeley National Laboratory, Leibniz Institut f\"ur Astrophysik Potsdam (AIP), Max-Planck-Institut f\"ur Astronomie (MPIA Heidelberg), Max-Planck-Institut f\"ur Astrophysik (MPA Garching), Max-Planck-Institut f\"ur Extraterrestrische Physik (MPE), National Astronomical Observatories of China, New Mexico State University, New York University, University of Notre Dame, Observat\'ario Nacional / MCTI, The Ohio State University, Pennsylvania State University, Shanghai Astronomical Observatory, United Kingdom Participation Group, Universidad Nacional Aut\'onoma de M\'exico, University of Arizona, University of Colorado Boulder, University of Oxford, University of Portsmouth, University of Utah, University of Virginia, University of Washington, University of Wisconsin, Vanderbilt University, and Yale University.

This publication makes use of data products from the Two Micron All Sky Survey, which is a joint project of the University of Massachusetts and the Infrared Processing and Analysis Center/California Institute of Technology, funded by the National Aeronautics and Space Administration and the National Science Foundation.

This work is based in part on data obtained as part of the UKIRT Infrared Deep Sky Survey.

This publication makes use of data products from the Wide-field Infrared Survey Explorer, which is a joint project of the University of California, Los Angeles, and the Jet Propulsion Laboratory/California Institute of Technology, funded by the National Aeronautics and Space Administration.

Numerical computations/simulations were carried out (in part) using the SuMIRe cluster operated by the Extragalactic
OIR group at ASIAA.

This work is supported by JSPS KAKENHI grant Nos. 18J01050 and 19K14759 (Y.T.), 17K05384 and 20H01946 (Y.U.), and 20H01949 (T.N.).

\vspace{5mm}
\facilities{Sloan, CTIO:2MASS, FLWO:2MASS, UKIRT, {\it WISE}}

\software{IDL, IDL Astronomy User's Library \citep{Landsman}, {\sf X-CIGALE} \citep{Boquien,Yang}, {\tt QSFit} \citep{Calderone}, {\tt TOPCAT} \citep{Taylor}.}

\appendix
\section{A value-added catalog for SDSS DR16 quasars at $z < 0.7$}
\label{app1}

We provide physical properties of 37,181 quasars at $z < 0.7$ selected from the SDSS DR16Q. 
The catalog description is summarized in Table \ref{catalog}.

\startlongtable
\begin{deluxetable}{lccl}
\tablecaption{Physical properties of 37,181 quasars at $z < 0.7$ selected from the SDSS DR16. \label{catalog}}
\tablehead{
\colhead{Column name} & \colhead{Format} & \colhead{Unit} & \colhead{Description}
}
\startdata
SpecObjID		& 	LONG	&			&	Unique id in the SDSS DR16 \\
Plate 			& 	INT32 	& 			& 	Spectroscopic plate number \\
MJD				& 	INT32 	&			&	Modified Julian day of the spectroscopic observation \\	
FiberID			& 	INT16 	&			&	Fiber ID number \\	
R.A.			&	DOUBLE	&	degree	& Right Assignation (J2000.0) from the SDSS DR16Q \\
Decl.			&	DOUBLE	&	degree	& Declination (J2000.0) from the SDSS DR16Q \\
Redshift		&	DOUBLE 	& 			& Redshift (that is taken from a column, ``$Z$'' in the SDSS DR16Q) \\
rechi2\_QSFIT	& DOUBLE	&			& Reduced $\chi^{2}$ derived from {\tt QSFIT} \\
Cont\_L3000		&	DOUBLE	&	erg s$^{-1}$	&	AGN continuum luminosity at the rest-frame 3000 \AA ~derived from {\tt QSFIT}\\
Cont\_L3000\_err&	DOUBLE	&	erg s$^{-1}$	&	Uncertainty of Cont\_L3000 derived from {\tt QSFIT} \\
Slope\_3000		&	DOUBLE	&					&	AGN continuum slope at the rest-frame 3000 \AA ~derived from {\tt QSFIT}\\
Slope\_3000\_err&	DOUBLE	&					&	Uncertainty of Slope\_3000 derived from {\tt QSFIT} \\
Cont\_3000\_Qual	&	INT32	&				&	Quality flag of continuum at the rest-frame 3000 \AA~derived from {\tt QSFIT} \\ 
					&			&				&	(see Appendix A (xix) in \citealt{Calderone}) \\
Cont\_L5100		&	DOUBLE	&	erg s$^{-1}$	&	AGN continuum luminosity at the rest-frame 5100 \AA ~derived from {\tt QSFIT}\\
Cont\_L5100\_err&	DOUBLE	&	erg s$^{-1}$	&	Uncertainty of Cont\_L5100 derived from {\tt QSFIT} \\
Slope\_5100		&	DOUBLE	&					&	AGN continuum slope at the rest-frame 5100 \AA ~derived from {\tt QSFIT}\\
Slope\_5100\_err	&	DOUBLE	&				&	Uncertainty of Slope\_5100 derived from {\tt QSFIT} \\
Cont\_5100\_Qual	&	INT32	&				&	Quality flag of continuum at the rest-frame 5100 \AA~derived from {\tt QSFIT} \\ 
					&			&				&	(see Appendix A (xix) in \citealt{Calderone}) \\
Lum\_MgII			&	DOUBLE	&erg s$^{-1}$	&	Line luminosity of Mg{\,\sc ii} derived from {\tt QSFIT}	\\
Lum\_MgII\_err		&	DOUBLE	&erg s$^{-1}$	&	Uncertainty of Lum\_MgII derived from {\tt QSFIT}  \\	
FWHM\_MgII			&	DOUBLE	&	km s$^{-1}$	&	FWHM of Mg{\,\sc ii} derived from {\tt QSFIT}	\\
FWHM\_MgII\_err		&	DOUBLE	&	km s$^{-1}$	&	Uncertainty of FWHM\_MgII derived from {\tt QSFIT}  \\
EW\_MgII			&	DOUBLE	&	\AA			&	Equivalent width of Mg{\,\sc ii} derived from {\tt QSFIT}	\\
EW\_MgII\_err		&	DOUBLE	&	\AA			&	Uncertainty of EW\_MgII derived from {\tt QSFIT}  \\
MgII\_Qual			&	INT32	&				&	Quality flag of Mg{\,\sc ii} line derived from {\tt QSFIT} \\
					&			&				&	(see Appendix A (xxxvi) in \citealt{Calderone}) \\	
Lum\_HB\_BR			&	DOUBLE	&erg s$^{-1}$	&	Line luminosity of H$\beta$ (broad component) derived from {\tt QSFIT}	\\
Lum\_HB\_BR\_err	&	DOUBLE	&erg s$^{-1}$	&	Uncertainty of Lum\_HB\_BR derived from {\tt QSFIT}  \\			
FWHM\_HB\_BR		&	DOUBLE	&	km s$^{-1}$	&	FWHM of H$\beta$ (broad component) derived from {\tt QSFIT}	\\
FWHM\_HB\_BR\_err	&	DOUBLE	&	km s$^{-1}$	&	Uncertainty of FWHM\_HB\_BR	derived from {\tt QSFIT}  \\
EW\_HB\_BR			&	DOUBLE	&	\AA			&	Equivalent width of H$\beta$ (broad component)  derived from {\tt QSFIT}	\\
EW\_HB\_BR\_err		&	DOUBLE	&	\AA			&	Uncertainty of EW\_HB\_BR derived from {\tt QSFIT}  \\
HB\_BR\_Qual	&	INT32	&					&	Quality flag of H$\beta$ (broad component) line derived from {\tt QSFIT} \\
					&			&				&	(see Appendix A (xxxvi) in \citealt{Calderone}) \\
Lum\_HB\_NA			&	DOUBLE	&erg s$^{-1}$	&	Line luminosity of H$\beta$ (narrow component) derived from {\tt QSFIT}	\\
Lum\_HB\_NA\_err	&	DOUBLE	&erg s$^{-1}$	&	Uncertainty of Lum\_HB\_NA derived from {\tt QSFIT}  \\	
FWHM\_HB\_NA		&	DOUBLE	&	km s$^{-1}$	&	FWHM of H$\beta$ (narrow component) derived from {\tt QSFIT}	\\
FWHM\_HB\_NA\_err	&	DOUBLE	&	km s$^{-1}$	&	Uncertainty of FWHM\_HB\_NA	derived from {\tt QSFIT}  \\
EW\_HB\_NA			&	DOUBLE	&	\AA			&	Equivalent width of H$\beta$ (narrow component)  derived from {\tt QSFIT}	\\
EW\_HB\_NA\_err		&	DOUBLE	&	\AA			&	Uncertainty of EW\_HB\_NA derived from {\tt QSFIT}  \\
HB\_NA\_Qual	&	INT32	&					&	Quality flag of H$\beta$ (narrow component) line derived from {\tt QSFIT} \\
					&			&				&	(see Appendix A (xxxvi) in \citealt{Calderone}) \\	
Lum\_OIII\_NA		&	DOUBLE	&erg s$^{-1}$	&	Line luminosity of [O{\,\sc iii}]$\lambda$5007  (narrow component) derived from {\tt QSFIT}	\\
Lum\_OIII\_NA\_err	&	DOUBLE	&erg s$^{-1}$	&	Uncertainty of Lum\_HB\_NA derived from {\tt QSFIT}  \\		
FWHM\_OIII\_NA		&	DOUBLE	&	km s$^{-1}$	&	FWHM of [O{\,\sc iii}]$\lambda$5007 (narrow component) derived from {\tt QSFIT}	\\
FWHM\_OIII\_NA\_err	&	DOUBLE	&	km s$^{-1}$	&	Uncertainty of FWHM\_OIII\_NA derived from {\tt QSFIT}  \\
EW\_OIII\_NA		&	DOUBLE	&	\AA			&	Equivalent width of [O{\,\sc iii}]$\lambda$5007 (narrow component)  derived from {\tt QSFIT}	\\
EW\_OIII\_NA\_err	&	DOUBLE	&	\AA			&	Uncertainty of EW\_OIII\_NA derived from {\tt QSFIT}  \\
OIII\_NA\_Qual		&	INT32	&				&	Quality flag of [O{\,\sc iii}]$\lambda$5007 (narrow component) line derived from {\tt QSFIT} \\
					&			&				&	(see Appendix A (xxxvi) in \citealt{Calderone}) \\	
Lum\_OIII\_BW		&	DOUBLE	&erg s$^{-1}$	&	Line luminosity of [O{\,\sc iii}]$\lambda$5007  (blue wing component) derived from {\tt QSFIT}	\\
Lum\_OIII\_BW\_err	&	DOUBLE	&erg s$^{-1}$	&	Uncertainty of Lum\_HB\_BW derived from {\tt QSFIT}  \\		
FWHM\_OIII\_BW		&	DOUBLE	&	km s$^{-1}$	&	FWHM of [O{\,\sc iii}]$\lambda$5007 (blue wing  component) derived from {\tt QSFIT}	\\
FWHM\_OIII\_BW\_err	&	DOUBLE	&	km s$^{-1}$	&	Uncertainty of FWHM\_OIII\_BW derived from {\tt QSFIT}  \\
EW\_OIII\_BW		&	DOUBLE	&	\AA			&	Equivalent width of [O{\,\sc iii}]$\lambda$5007 (blue wing  component)  derived from {\tt QSFIT}	\\
EW\_OIII\_BW\_err	&	DOUBLE	&	\AA			&	Uncertainty of EW\_OIII\_BW derived from {\tt QSFIT}  \\
Voff\_OIII\_BW		&	DOUBLE	&	km s$^{-1}$	&	Velocity offset of [O{\,\sc iii}]$\lambda$5007 (blue wing  component) derived from {\tt QSFIT}	\\
Voff\_OIII\_BW\_err	&	DOUBLE	&	km s$^{-1}$	&	Uncertainty of Voff\_OIII\_BW derived from {\tt QSFIT}  \\
OIII\_BW\_Qual		&	INT32	&				&	Quality flag of [O{\,\sc iii}]$\lambda$5007 (blue wing  component) line derived from {\tt QSFIT} \\
					&			&				&	(see Appendix A (xxxvi) in \citealt{Calderone}) \\		
Lum\_HA\_BR			&	DOUBLE	&erg s$^{-1}$	&	Line luminosity of H$\beta$ (broad component) derived from {\tt QSFIT}	\\
Lum\_HA\_BR\_err	&	DOUBLE	&erg s$^{-1}$	&	Uncertainty of Lum\_HA\_BR derived from {\tt QSFIT}  \\			
FWHM\_HA\_BR		&	DOUBLE	&	km s$^{-1}$	&	FWHM of H$\alpha$ (broad component) derived from {\tt QSFIT}	\\
FWHM\_HA\_BR\_err	&	DOUBLE	&	km s$^{-1}$	&	Uncertainty of FWHM\_HA\_BR	derived from {\tt QSFIT}  \\
EW\_HA\_BR			&	DOUBLE	&	\AA			&	Equivalent width of H$\alpha$ (broad component)  derived from {\tt QSFIT}	\\
EW\_HA\_BR\_err		&	DOUBLE	&	\AA			&	Uncertainty of EW\_HA\_BR derived from {\tt QSFIT}  \\	
HA\_BR\_Qual	&	INT32	&					&	Quality flag of H$\alpha$ (broad component) line derived from {\tt QSFIT} \\
					&			&				&	(see Appendix A (xxxvi) in \citealt{Calderone}) \\
Lum\_HA\_NA			&	DOUBLE	&erg s$^{-1}$	&	Line luminosity of H$\beta$ (narrow component) derived from {\tt QSFIT}	\\
Lum\_HA\_NA\_err	&	DOUBLE	&erg s$^{-1}$	&	Uncertainty of Lum\_HA\_NA derived from {\tt QSFIT}  \\			
FWHM\_HA\_NA		&	DOUBLE	&	km s$^{-1}$	&	FWHM of H$\alpha$ (narrow component) derived from {\tt QSFIT}	\\
FWHM\_HA\_NA\_err	&	DOUBLE	&	km s$^{-1}$	&	Uncertainty of FWHM\_HA\_NA	derived from {\tt QSFIT}  \\
EW\_HA\_NA			&	DOUBLE	&	\AA			&	Equivalent width of H$\alpha$ (narrow component)  derived from {\tt QSFIT}	\\
EW\_HA\_NA\_err		&	DOUBLE	&	\AA			&	Uncertainty of EW\_HA\_BR derived from {\tt QSFIT}  \\	
HA\_NA\_Qual	&	INT32	&					&	Quality flag of H$\alpha$ (narrow component) line derived from {\tt QSFIT} \\
					&			&				&	(see Appendix A (xxxvi) in \citealt{Calderone}) \\	
log\_MBH			& DOUBLE	&	$M_{\sun}$		&	Black hole mass (see Equation \ref{Eq}) \\
log\_MBH\_err		& DOUBLE	&	$M_{\sun}$		&	Uncertainty of log\_MBH \\
log\_Lbol			& DOUBLE	&	erg s$^{-1}$	&	Bolometric luminosity (see Section \ref{qsfit}) \\
log\_Lbol\_err		& DOUBLE	&	erg s$^{-1}$	&	Uncertainty of log\_Lbol \\
log\_lambda\_Edd 			& DOUBLE 	& 					&	Eddington ratio  \\
log\_lambda\_Edd\_err 		& DOUBLE 	& 					&	Uncertainty of log\_lambda\_Edd \\
rechi2\_XCIGALE		& DOUBLE	&	& Reduced $\chi^{2}$ derived from {\tt X-CIGALE}\\
Delta\_BIC			& DOUBLE	&	& BIC$_{\rm wopolar}$ -- BIC$_{\rm wpolar}$ (see Section \ref{BIC}) \\
E\_BV				& DOUBLE	&	&	Color excess ($E(B-V)$) derived from {\tt X-CIGALE} \\
E\_BV\_err			& DOUBLE	&	&	Uncertainty of color excess ($E(B-V)$) derived from {\tt X-CIGALE} \\
log\_M				& DOUBLE & $M_{\sun}$ & Stellar mass derived from {\tt X-CIGALE} \\
log\_M\_err 		& DOUBLE & $M_{\sun}$ & Uncertainty of stellar mass derived from {\tt X-CIGALE} \\
log\_SFR			& DOUBLE & $M_{\sun}$ yr$^{-1}$ & SFR derived from {\tt X-CIGALE} \\
log\_SFR\_err 		& DOUBLE & $M_{\sun}$ yr$^{-1}$ & Uncertainty of SFR derived from {\tt X-CIGALE} \\
log\_LIR			& DOUBLE & $L_{\sun}$ & IR luminosity derived from {\tt X-CIGALE} \\
log\_LIR\_err		& DOUBLE & $L_{\sun}$ & Uncertainty of IR luminosity derived from {\tt X-CIGALE} \\
log\_LIR\_AGN		& DOUBLE & $L_{\sun}$ & IR luminosity contributed from AGN derived from {\tt X-CIGALE} \\
log\_LIR\_AGN\_err	& DOUBLE & $L_{\sun}$ & Uncertainty of log\_LIR\_AGN derived from {\tt CIGALE} \\
log\_LIR\_AGN\_polar		& DOUBLE & erg s$^{-1}$	& IR luminosity contributed from AGN polar dust component \\
							&		 &			  	& derived from {\tt CIGALE} \\
log\_LIR\_AGN\_polar\_err	& DOUBLE & erg s$^{-1}$	& Uncertainty of log\_LIR\_polar derived from {\tt CIGALE} \\
log\_LIR\_AGN\_torus		& DOUBLE & erg s$^{-1}$	& IR luminosity contributed from AGN dust torus component \\
							&		 &			  	& derived from {\tt CIGALE} \\
log\_LIR\_AGN\_torus\_err	& DOUBLE & erg s$^{-1}$	& Uncertainty of log\_LIR\_torus derived from {\tt CIGALE} \\
CF\_AGN\_torus				& DOUBLE &				& Covering factor of AGN dust torus (see Equation \ref{Eq_CF}) \\
CF\_AGN\_torus\_err			& DOUBLE &				& Uncertainty of CF\_AGN\_torus	 \\
\enddata
\tablecomments{5720 quasars with ($\chi^2$/dof)$_{\rm QSFit} < 3.0$, ($\chi^2$/dof)$_{\rm CIGALE} < 3.0$, $\Delta$BIC $>$ 6, {\tt Cont\_5100\_Qual} = 0, and {\tt HB\_BR\_Qual} = 0 are used for a correlation analysis (see Section \ref{CF}). This table is available in its entirety in a machine-readable form in the online journal.}
\end{deluxetable}


\end{document}